\documentclass[a4paper,UKenglish,cleveref, autoref, thm-restate,authorcolumns]{lipics-v2019}
\usepackage[utf8]{inputenc}
\usepackage{babel}
\usepackage{hyperref}
\usepackage{amsthm,thmtools}
\usepackage{algorithm}
\usepackage{algpseudocode}
\usepackage{tikz}
\tikzstyle{vertex}=[circle, draw, inner sep=0pt, minimum size=5pt]

\usetikzlibrary{decorations.markings}
\usetikzlibrary{decorations.pathreplacing}
\usetikzlibrary{arrows.meta}
\usepackage{anyfontsize}
\usepackage{graphicx}
\usepackage{hyperref}
\usepackage{amsmath}
\usepackage{amsfonts}
\usepackage{xcolor}
\usepackage{amssymb}
\usepackage{amsmath}
\newcommand{\boundellipse}[3]
{(#1) ellipse (#2 and #3)
}
\usetikzlibrary{shapes.geometric}
\tikzstyle{square}=[draw, shape=regular polygon, regular polygon sides=4,draw,inner sep=0pt,minimum
size=0.225cm]
\tikzstyle{triangle}=[draw, shape=regular polygon, regular polygon sides=3,draw,inner sep=0pt,minimum
size=0.3cm]

\definecolor{azure}{rgb}{0.0, 0.5, 1.0}
\definecolor{pink}{rgb}{0.84, 0.09, 0.41}
\definecolor{magenta}{rgb}{0.8, 0.0, 0.8}
\definecolor{cyan}{rgb}{0.0, 1.0, 1.0}
\definecolor{green1}{rgb}{0, 1, 0}
\definecolor{green}{rgb}{0, 1, 0}
\definecolor{brown}{rgb}{0.65, 0.16, 0.16}
\definecolor{aquamarine}{rgb}{0.5, 1.0, 0.83}
\definecolor{battleshipgrey}{rgb}{0.52, 0.52, 0.51}
\definecolor{cadetgrey}{rgb}{0.57, 0.64, 0.69}
\newtheorem{observation}{\bf Observation}


\title{Parameterized Algorithms for Locally Minimal Defensive Alliance} 

\titlerunning{Locally Minimal Defensive Alliance} 

\author{Ajinkya Gaikwad}{Indian Institute of Science Education and Research, Pune, India}{ ajinkya.gaikwad@students.iiserpune.ac.in}{}{}
\author{Soumen Maity}
{Indian Institute of Science Education and Research, Pune, India}{soumen@iiserpune.ac.in}{}{}
\author{Saket Saurabh}{
The Institute of Mathematical Sciences, Chennai, India \and University of Bergen, Bergen, Norway} {saket@imsc.res.in}{}{}

\authorrunning{A. Gaikwad, S. Maity and S. Saurabh} 

\Copyright{Ajinkya Gaikwad} 
\ccsdesc[100]{General and reference~General literature}
\ccsdesc[100]{General and reference}

\keywords{ Parameterized Complexity, FPT,  W[1]-hard, treewidth}
\category{} 

\relatedversion{} 

\supplement{}

\nolinenumbers

\EventEditors{John Q. Open and Joan R. Access}
\EventNoEds{2}
\EventLongTitle{42nd Conference on Very Important Topics (CVIT 2016)}
\EventShortTitle{}
\EventAcronym{CVIT}
\EventYear{2016}
\EventDate{December 24--27, 2016}
\EventLocation{Little Whinging, United Kingdom}
\EventLogo{}
\SeriesVolume{42}
\ArticleNo{23}

\begin{document}

\maketitle

\begin{abstract}
A set $D$ of vertices of a graph is a \emph{defensive alliance} if, 
for each element of $D$, 
the majority of its neighbours are in $D$.
We consider the notion of local minimality in this paper. We are interested in finding  a locally 
 minimal defensive alliance of maximum size.
  In {\sc Locally Minimal Defensive Alliance} problem, given an undirected graph $G$, a positive integer $k$, the question is to check whether $G$ has a locally minimal defensive alliance of size at least $k$. 
 This problem is known to be NP-hard  but its parameterized complexity remains open 
until now.  We enhance our understanding of the problem from the
viewpoint of parameterized complexity.
The main results of the paper are the following: (1) {\sc Locally Minimal Defensive Alliance} restricted to the graphs of minimum degree at least 2 is fixed-parameter tractable (FPT) when parameterized by the combined parameters solution size $k$, and maximum degree $\Delta$ of the input graph,
(2) 
{\sc Locally Minimal Defensive Alliance}  on the graphs of minimum degree at least 2, admits
a kernel with at most $k^{k^{\mathcal{O}(k)}}$ vertices. 
In particular, the problem 
parameterized by $k$ restricted to  $C_3$-free and $C_4$-free graphs 
of minimum degree at least 2, admits
a kernel with at most $k^{\mathcal{O}(k)}$ vertices.
Moreover, we prove that the problem on planar graphs of minimum degree at least 2, admits an FPT algorithm with running time $\mathcal{O}^{*}(k^{2^{\mathcal{O}(\sqrt{k})}})$.  Finally we  prove that (4) {\sc Locally Minimal Defensive Alliance Extension} is NP-complete.

\end{abstract}

\keywords{Parameterized Complexity \and FPT  \and locally minimal defensive alliance }

\section{Introduction}
Throughout history, humans have formed communities, guilds, faiths etc in the hope of coming together with a group of people having similar requirements, visions and goals. Their reasons to do so, usually rest on the fact that any group with common interests often provides added mutual benefits to the union in fields of trade, culture, defense, etc as compared to the individual. Such activities are commonly seen in the present day, in areas of geo-politics, cultures, trades, economics, unions etc and are popularly termed as \emph{alliances}. 
Based on the structure, formation and goals of an alliance, many variations of the problem exist in graph theory. 
A defensive alliance is usually formed with the aim of defending its members against non-members, and hence it is natural to ask that each member of the alliance should have more friends within the alliance (including oneself) than outside.
 Similarly, an offensive alliance is formed with the inverse goal of offending or attacking non-members of the alliance. 
 It is known that the problems of finding small 
defensive and offensive alliances are NP-complete. We enhance our understanding of the problems from the 
viewpoint of parameterized complexity.
  \emph{Strong} versions of the above problems do not consider the self to be a friend, and the \emph{minimal} versions try to find those alliances which lose the required property in the absence of any member.
 In 2004, Kristiansen, Hedetniemi, and Hedetniemi \cite{kris}
introduced defensive, offensive, and powerful alliances, and further studied by Shafique \cite{HassanShafique2004PartitioningAG}
and other authors \cite{small,BAZGAN2019111,Cami2006OnTC,SIGARRETA20061345,ROD,SIGARRETA20091687,SIGA,Enciso2009AlliancesIG,BLIEM2018334,Fernau,FERNAU2009177,Lindsay}. 
The theory of alliances in graphs has been studied 
intensively \cite{frick,Cami2006OnTC,10.5614/ejgta.2014.2.1.7} both from a combinatorial and from a computational perspective. 
As mentioned in \cite{BAZGAN2019111}, the 
focus has been mostly on finding small alliances, although studying large
alliances does not only make a lot of sense from the original motivation of these notions, 
but was actually also delineated in the very first papers on alliances \cite{kris}.

Note that being a defensive alliance is not a hereditary property, that is, a superset or subset of a
defensive alliance is not necessarily a  defensive alliance.  Shafique \cite{HassanShafique2004PartitioningAG} called 
an alliance a {\it locally minimal alliance} if the set obtained by removing any vertex of the
alliance is not an alliance. Bazgan  et al. \cite{BAZGAN2019111} considered another notion 
of alliance that they called a {\it globally minimal alliance} which has the property that 
no proper subset is an alliance. In this paper we are interested in finding locally minimal 
alliances of size at least $k$.  Bazgan et al. \cite{BAZGAN2019111} proved that deciding if a graph contains a  locally minimal defensive alliance of size at least $k$ is NP-complete, even when restricted to 
 bipartite graphs with average degree less than 5.6.  Clearly, the motivation is that big communities where every member
still matters somehow are of more interest than really small communities. Also, there is a 
general mathematical interest in such type of problems, see \cite{Manlove1998MinimaximalAM}.
\par Parameterized complexity of alliance problems is also well studied with both natural and structural parameterizations.
When it comes to structural parameterization, the alliance problems such as defensive and offensive alliances are shown to be W[1]-hard when parameterized by vertex deletion set into trees of constant height, see \cite{10.1007/978-3-030-92681-6_45,GAIKWAD2022136}. It is not known either if above problems admits FPT algorithm when parameterized by other structural parameters such as modular width and cluster vertex deletion set.
Fortunately, the same problems admit FPT algorithm when parameterized by solution size, see \cite{DA-FPT}.
As seen in \cite{DBPL-LMDA,GAIKWAD2022106253}, the problem of finding locally and globally minimal alliances of maximum size as well shows the same trend of being intractable when parameterized by structural parameters such as treewidth. In particular, the paper \cite{DBPL-LMDA} showed that {\sc Exact Connected Locally Minimal Defensive Alliance} parameterized by treewidth is W[1]-hard.
Therefore it is an interesting question to know whether problem of finding minimal alliance of size at least $k$ admits FPT algorithm when parameterized by solution size.

\section{Definitions and Preliminaries}
Throughout this article, $G=(V,E)$ denotes a finite, simple and undirected graph of order $|V|=n$.
The {\it (open) neighbourhood} $N_G(v)$ of a vertex 
$v\in V(G)$ is the set $\{u~|~(u,v)\in E(G)\}$. The {\it closed neighbourhood} $N_G[v]$ of a vertex $v\in V(G)$ is the set
$\{v\} \cup N_G(v)$.  The {\it degree} of $v\in V(G)$ is $|N_G(v)|$ and denoted by $d_G(v)$. The subgraph induced by 
$D\subseteq V(G)$ is denoted by $G[D]$. We use $d_D(v)$ to denote the number of neighbours of vertex $v$ in $D$. 
The complement of the vertex set $D$ in $V$ is denoted by $D^c$. The minimum degree of graph $G$ is denoted by $\delta(G)$.
\begin{definition}\rm
A non-empty set $D\subseteq V$ is a \emph{defensive alliance} in $G$ if for each $v\in D$, 
  $d_D(v)+1\geq d_{D^c}(v)$.  
\end{definition}
A vertex $v\in D$ is said to be \emph{protected} if $d_D(v)+1\geq d_{D^c}(v)$. Here $v$ has 
$d_D(v)+1$ defenders and $d_{D^c}(v)$ attackers in $G$.  
A set $D\subseteq V$ is a defensive alliance if every vertex
in $D$ is protected. 
\begin{definition}\rm
A vertex $v\in D$ is said to be {\it marginally protected} 
if it becomes unprotected when any of its neighbour in $D$ is moved from $D$ to $V\setminus D$.
A vertex $v\in D$ is said to be {\it overprotected} if it remains protected
even when any of its neighbour is moved from $D$ to $V\setminus D$. 
\end{definition}

\begin{definition}\rm \cite{BAZGAN2019111}
An alliance $D$ is called a {\it locally minimal alliance}
if for any $v\in D$, $D\setminus \{v\}$ is not an alliance.
\end{definition}

\noindent It is important to note  that if $D$ is a locally minimal  defensive alliance, 
 then for every vertex $v \in D$, at least one of its neighbours in $D$ is marginally protected.

 \begin{definition}\rm \cite{BAZGAN2019111}
An alliance $D$ is a {\it globally minimal alliance} or  shorter {\it minimal alliance} if no proper 
subset is an alliance.
\end{definition}
\begin{figure}[!h]
   \centering
   \begin{tikzpicture}[scale=0.5]
\node[vertex](v1)at(6,0){$1$};
\node[vertex](v2)at(0,-2){2};
\node[vertex](v3)at(1.5,-2){3};
\node[vertex](v4)at(3,-2){4};
\node[vertex](v5)at(4.5,-2){5};
\node[vertex](v6)at(6,-2){6};
\node[vertex](v7)at(7.5,-2){7};
\node[vertex](v8)at(9,-2){8};
\node[vertex](v9)at(10.5,-2){9};
\path
    (v1) edge (v2)
    (v1) edge (v3)
    (v1) edge (v4)
    (v1) edge (v5)
    (v1) edge (v6)
    (v1) edge (v7)
    (v1) edge (v8)
    (v1) edge (v9)
    (v3) edge (v2)
    (v4) edge (v3)
    (v5) edge (v4)
    (v6) edge (v5)
    (v7) edge (v6)
    (v8) edge (v7)
    (v9) edge (v8);
   \end{tikzpicture}
\caption{The set $D_1=\{2,3,5,6,8,9\}$  is a locally minimal defensive alliance of size 6 and  $D_2=\{1,2,4,6,8\}$ is a globally minimal defensive alliance of size 5 in $G$.}
  \label{treeLMDA}
 \end{figure}
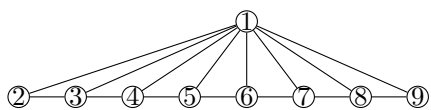

A  defensive alliance $D$ is connected if the subgraph induced by 
$D$ is connected. An alliance $D$ is called a {\it  connected locally minimal alliance}
if for any $v\in D$, $D\setminus \{v\}$ is not a connected alliance.  Notice that
any globally minimal alliance is also connected. 
In this paper, we consider {\sc Locally Minimal Defensive Alliance} and {\sc  Connected Locally Minimal Defensive Alliance}.
 We define the problems as follows:
    \vspace{3mm}
    \\
    \fbox
    {\begin{minipage}{38.7em}\label{FFVS0 }
       {\sc Locally Minimal Defensive Alliance}\\
        \noindent{\bf Input:} An undirected graph $G=(V,E)$ and an  integer $k$.
    
        \noindent{\bf Question:} Does $G$ have a locally minimal 
     defensive alliance $D$ with $|D|\geq k$?
        
    \end{minipage} }
  \vspace{3mm}

\noindent We now recall the definition of height of a tree and diameter of a connected graph.  
 \begin{definition}\rm The \emph{level} of a vertex $v$ in a rooted tree is the number of edges along the unique path between it and the root node.
 The \emph{height} of a rooted tree is the maximum of the levels of vertices. 
 \end{definition}
 Let $G=(V,E)$ be a connected graph. For any $u,v\in V$, let $d(u,v)$ denote the distance between $u$ and $v$, that is, the length of a shortest path between $u$ and $v$ in $G$. The \emph{diameter} of $G$ is defined as
 $$diam(G)=\max\limits_{u,v\in V} d(u,v).$$ Note that if $diam(G)\geq 2k$, then the   BFS tree height is at least
$k$.\\

A \emph{parameterized problem} is a language $L\subseteq \Sigma^{\star} \times \mathbb{N}$,
where $\Sigma $
is a fixed, finite alphabet. For an instance $(x,k)\in \Sigma^{\star} \times \mathbb{N}$,
$k$ is called the \emph{parameter}. A parameterized problem $\mathcal{P}$ is \emph{fixed-parameter tractable} (FPT in short) if a given instance $(x,k)$ can be solved in time $f(k)\cdot {|(x,k)|}^c$
 where $f$ is some (usually computable) function, and $c$ is a constant. Parameterized complexity classes are defined with respect to {\it fpt-reducibility}. A parameterized problem $\mathcal{P}$ is {\it fpt-reducible} to 
 $\mathcal{Q}$ if in time  $f(k) \cdot {|(x,k)|}^c$, one can transform an instance 
 $(x, k)$ of $\mathcal{P}$ into an instance $(x', k')$ of $Q$ such that $(x, k) \in \mathcal{P}$
 if and only if  $(x',k') \in \mathcal{Q}$, and $k'\leq g(k)$, where $f$ and $g$ are computable functions depending only on $k$. Owing to the definition, if $\mathcal{P}$ {\it fpt-reduces} to $\mathcal{Q}$ and $\mathcal{Q}$ is fixed-parameter tractable then $\mathcal{P}$ is fixed-parameter tractable as well.

What makes the theory more interesting is a hierarchy of intractable parameterized problem classes above FPT which helps in distinguishing those problems that are not fixed parameter tractable. 
Central to parameterized complexity is the following hierarchy of complexity classes, defined by the closure of canonical problems under {\it fpt-reductions}: FPT $\subseteq$ W[1] $\subseteq$ W[2] $\subseteq \ldots \subseteq $ XP. All inclusions are believed to be strict. In particular, FPT $\neq$  W[1] under the Exponential Time Hypothesis \cite{Paturi}.
The class W[1] is the analog of NP in parameterized complexity. A major goal in parameterized complexity is to distinguish between parameterized problems which are in FPT
 and those which are \emph{W[1]-hard}, i.e., those to which every problem in W[1] is \emph{fpt-reducible}. There are many problems shown to be complete for W[1], or equivalently \emph{W[1]-complete}, including the {\sc MultiColored Clique} (MCC) problem \cite{Downey}.

 Closely related to fixed-parameter tractability is the notion of preprocessing. 
 A \emph{preprocessing algorithm} $\mathcal{A}$ takes as input an instance $(x,k)\in \Sigma^{\star} \times \mathbb{N}$ of $\mathcal{P}$, works in polynomial time, and returns
 an equivalent instance $(x',k')$ of $\mathcal{P}$. The \emph{output size} of a 
 preprocessing algorithm $\mathcal{A}$ is a function $\mbox{size}_{\mathcal{A}}:
 \mathbb{N}\rightarrow \mathbb{N}\cup\{\infty\}$ defined as follows:
 $$\mbox{size}_{\mathcal{A}}(k)=\sup \{ |x'|+k'~:~(x',k')=\mathcal{A}(x,k),~x\in \Sigma^*\}.$$ \emph{Kernelization algorithms} are exactly these 
 preprocessing algorithms whose output size is finite and bounded by a computable function of the parameter $k$. 
It is easy to show that a parameterized problem is in FPT if and only if there is a kernelization algorithm. 
A \emph{polynomial kernel} is a kernel, whose size can be bounded by a polynomial in the parameter.
 We refer to \cite{marekcygan,Downey} for further details on parameterized complexity. \\

\noindent Our results are as follows:
\begin{itemize}
 \item  {\sc Locally Minimal DA Extension} is NP-complete. 
 \item {\sc Locally Minimal Defensive Alliance} is FPT when parameterized by the combined parameters solution size and maximum degree of the input graph.
 \item {\sc Locally Minimal Defensive Alliance}  on the graphs 
 of minimum degree at least 2, admits
a kernel with at most $k^{k^{\mathcal{O}(k)}}$ vertices. 
In particular, we prove that the problem on $C_3$-free or $C_4$-free graphs of minimum degree at least 2, admits a kernel with at most $k^{\mathcal{O}(k)}$ vertices.
\item {\sc Locally Minimal Defensive Alliance} on the planar graphs of minimum degree at least 2, admits an FPT algorithm with running time  $\mathcal{O}^{*}( k^{2^{\mathcal{O}(\sqrt{k})}})$.
\end{itemize}    
 
 \subsection{On {\sc Locally Minimal Defensive Alliance Extension}}
We demonstrate that constructing even a XP algorithm for {\sc Locally Minimal Defensive Alliance} is not that straightforward.
The brute-force approach to find a locally minimal defensive alliance of size at least $k$ is to consider 
all $\dbinom{n}{k}$ subsets of size $k$ and then check for each one if there exists a locally minimal defensive alliance containing it.
We show that given an arbitrary set $S\subseteq V$, it is NP-complete to determine whether there exists a locally minimal defensive alliance containing $S$.
 Formally, we want to discuss the following problem:

\vspace{3mm}
 \noindent   \fbox
    {\begin{minipage}{38.7em}\label{FFVS }
       {\sc Locally Minimal DA Extension}\\
        \noindent{\bf Input:} A graph $G=(V,E)$, a set $S\subseteq V$.\\
     \noindent{\bf Question:}  Does $G$ have a locally minimal 
     defensive alliance $D$ with $S\subseteq D$?
    \end{minipage} }
    \vspace{3mm}
    \\
Let us also discuss the case where $S$ contains only one vertex. Note that given a vertex $v$, there always exists a defensive alliance containing it. 
But such a statement does not hold when it comes to locally minimal defensive 
alliance, that is, given a vertex $v$ there may or may not exist a locally minimal defensive alliance containing $v$.
For example, consider the star graph $S_{n}$ with $v\in V(S_{n})$ such that $d(v)=n-1$.
We observe that there is no locally minimal defensive alliance containing $v$. It is not clear if this result holds when $|S|=1$. 
The example of a star graph may be a very special case. 
We do not know of any graph $G$ with minimum degree two, where given a vertex $v$ there is 
no locally minimal defensive alliance containing $v$.  
It may be possible that such an example does not even exist. We have discussed more about the importance of the existence of such an example in the conclusion.
We now show that {\sc Locally Minimal DA Extension} is NP-hard. 
\begin{theorem}\label{thm:NPlmda}
{\sc Locally Minimal DA Extension} is NP-complete. 
\end{theorem}

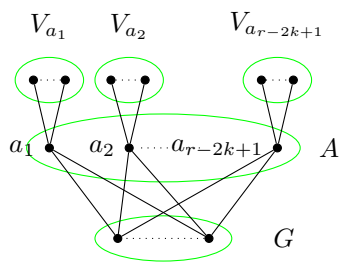
\begin{figure}[ht]
    \centering
 \begin{tikzpicture}[scale=0.6]
 \draw[draw=green] (0,0) ellipse (3cm and .75cm);
 
 \node (A) at (3,0) [label=right:$A$] {};
 
 \draw[draw=green] (0,-2) ellipse (1.5cm and .5cm);
 
 \node (A1) at (2,-2) [label=right:$G$] {};
 
 \node (A1) at (-2.5,2) [label=above:$V_{a_{1}}$] {};
 \node (A2) at (-.75,2) [label=above:$V_{a_{2}}$] {};
  \node (A3) at (2.5,2) [label=above:$V_{a_{r-2k+1}}$] {};
 
 \draw[draw=green] (-2.5,1.5) ellipse (.75cm and .5cm);
 \draw[draw=green] (-.75,1.5) ellipse (.75cm and .5cm);
 \draw[draw=green] (2.5,1.5) ellipse (.75cm and .5cm);
 
\node[circle, draw, inner sep=0pt, fill=black,minimum size=3pt] (a1) at (-2.5,0) [label=left:$a_{1}$] {};

\node[circle, draw, inner sep=0pt, fill=black,minimum size=3pt] (a11) at (-2.15,1.5) [label=left:] {};
\node[circle, draw, inner sep=0pt, fill=black,minimum size=3pt] (a12) at (-2.85,1.5) [label=left:] {};

\node[circle, draw, inner sep=0pt, fill=black,minimum size=3pt] (a2) at (-.75,0) [label=left:$a_{2}$] {};

\node[circle, draw, inner sep=0pt, fill=black,minimum size=3pt] (a21) at (-1.15,1.5) [label=left:] {};
\node[circle, draw, inner sep=0pt, fill=black,minimum size=3pt] (a22) at (-.4,1.5) [label=left:] {};

\node[circle, draw, inner sep=0pt, fill=black,minimum size=3pt] (a3) at (2.5,0) [label=left:$a_{r-2k+1}$] {};

\node[circle, draw, inner sep=0pt, fill=black,minimum size=3pt] (a31) at (2.15,1.5) [label=left:] {};
\node[circle, draw, inner sep=0pt, fill=black,minimum size=3pt] (a32) at (2.85,1.5) [label=left:] {};
 
\node[circle, draw, inner sep=0pt, fill=black,minimum size=3pt] (g1) at (1,-2) [label=left:] {};
\node[circle, draw, inner sep=0pt, fill=black,minimum size=3pt] (g2) at (-1,-2) [label=left:] {};

\draw[dotted](a11)--(a12);
\draw[dotted](a21)--(a22);
\draw[dotted](a31)--(a32);

\draw[dotted](0.8,0)--(-.5,0);

\draw[dotted](g1)--(g2);

\draw(a1)--(a11);
\draw(a1)--(a12);
\draw(a2)--(a21);
\draw(a2)--(a22);
\draw(a3)--(a31);
\draw(a3)--(a32);
\draw(g1)--(a1);
\draw(g2)--(a1);
\draw(g1)--(a2);
\draw(g2)--(a2);
\draw(g1)--(a3);
\draw(g2)--(a3);
 
 \end{tikzpicture}
    \caption{The reduction of
an instance $G$ of {\sc Clique}  on regular graphs to an instance
$G'$ of {\sc Locally Minimal DA Extension}  in Theorem \ref{thm:NPlmda}.}
    \label{Ext}
\end{figure}

\proof It is easy to see that  {\sc Locally Minimal DA Extension} is in NP. 
NP-hardness can be shown by reduction  from {\sc Clique} on regular graphs. 
 Let $I=(G,k)$ be an instance of {\sc Clique}, where $G$ is an $r$-regular graph.  We construct an instance $I'=(G',S)$ of {\sc Locally Minimal DA Extension} as follows.
 See Figure \ref{Ext}.
 The construction 
of $G'$ starts with $G':=G$ and then add some new vertices and edges.
 First, we introduce a set $A=\{a_{1},\ldots,a_{r-2k+1}\}$ of $r-2k+1$ new vertices
 into $G'$.
 For every vertex $a_i \in A$, we introduce a set $V_{a_{i}}$ of $n-2k-1$ vertices into 
 $G'$ and make them adjacent to $a_i$. We make every vertex of $V(G)$ adjacent to 
 every vertex of $A$. This completes the construction of  $G'$. 
 Finally we set $S= \bigcup\limits_{i=1}^{r-2k+1} V_{a_{i}}$. 
 To prove the correctness of the reduction, we claim that $G$ has a $k$-clique if and only if $G'$ admits a locally minimal defensive alliance 
 containing $S$. Assume first that $G$ has a clique $C$ of size $k$.
 We claim that $D = S \cup A \cup C$ is a locally minimal defensive alliance in $G'$. 
 Let $x$ be an arbitrary element of $D$. If $x$ is an element of $S$, then  the only neighbor of $x$ in $G'$ is in $D$,
so $x$ can trivially defend itself. If $x$ is an element of $A$, then 
the neighbours of $x$ in $D$ are the elements of $V_{x} \cup C$. Thus
including itself $x$ has $(n-2k-1)+k+1=n-k$ neighbours inside $D$ and $n-k$ neighbours outside $D$, so $x$ is marginally protected.
If $x$ is an element of  $C$ then, including itself, $x$ has $(r-2k+1)+k=r-k+1$ neighbours inside $D$ and $r-k+1$ neighbours outside $D$ as $G$ is a $r$-regular graph. 
  Also every vertex in $D$ has a marginally protected neighbour. This proves that $D$ is a locally minimal defensive alliance containing $S$. 
\par To prove the reverse direction of the equivalence, suppose $G'$ has a locally minimal defensive alliance $D$ containing $S$.
Since every vertex in $S$ is of degree one, its only neighbour must be contained in locally minimal defensive alliance and it must be marginally protected. That means, 
the vertices of $A$ are in $D$ and every vertex $a$ of $A$ must be marginally protected. 
Note that, including itself, $a$ has total $2n-2k$ neighbours in $G'$. As $a$ is marginally 
protected, 
$n-k$ neighbours are inside $D$ and 
the remaining $n-k$ neighbours are outside $D$. 
It implies that $|D\cap V(G)|=k$ as every vertex of  $A$ is adjacent to 
every vertex of  $V(G)$.
Let $D\cap V(G)=C$. We will show that $C$ is a clique. 
Let $u$ be an arbitrary vertex in $C$, then $d_{G'}(u)=2r-2k+1$
as $u$ is adjacent to $r-2k+1$ elements of $A$ and $r$ elements of $G$. 
Clearly, including itself, $u$ must have at least $r-k+1$ neighbours inside $D$.
The neighbors of $u$ that are also in $D$ consist of the elements of $A$ and at least $k-1$ elements of $C$.
Since $|C|=k$, $u$ must be adjacent to every vertex of $C$.
Therefore $C$ is a  clique. This completes the proof of Theorem \ref{thm:NPlmda}. \qed 

\section{FPT algorithm for {\sc Locally Minimal Defensive Alliance} parameterized by $k+\Delta$}

In this section we present  FPT algorithms for {\sc Locally Minimal Defensive Alliance} and {\sc Connected Locally Minimal Defensive Alliance} when parameterized by the combined parameters solution size $k$, and maximum degree $\Delta$ of the input graph. In particular, we will prove the following theorems.

\begin{theorem}\label{thm:lmda}
{\sc Locally Minimal Defensive Alliance} on the graphs of minimum degree at least 2,  admits a kernel of size ${\Delta^{\mathcal{O}(k)}}$, when parameterized by  $k+\Delta$.
\end{theorem}

\begin{theorem}\label{thm:clmda}
{\sc Connected Locally Minimal Defensive Alliance} on the graphs of minimum degree at least 2, admits an FPT algorithm when 
parameterized by $k+\Delta$.
\end{theorem}

\noindent In order to prove Theorem \ref{thm:lmda}, we need the following simple observations and lemmas. From now on, we  consider graphs of minimum degree at least $2$; so any defensive alliance in $G$ must have size at least 2. 

\begin{observation}\label{obs1}\rm Let $G$ be a graph of minimum degree $2$ and $D$ be a defensive alliance of
size at least $2$ in $G$. 
Then $D$  contains a locally minimal defensive alliance of size at least $2$ and such a locally minimal defensive alliance can be found in polynomial time.
\end{observation}
Pick an arbitrary vertex $u\in D$ and check if $D\setminus \{u\}$ is a defensive alliance. If yes, then remove $u$ otherwise do not remove it. We repeat this until no vertex can be removed;
at the end the defensive alliance we get is a locally minimal defensive alliance. Also observe that it is of size at least 2 as $\delta(G) \geq 2$. 

\begin{observation}\label{Union1}\rm 
Suppose  $G$ with $\delta(G)\geq 2$ has  $p$ locally minimal defensive alliances 
$S_{1},\ldots,S_{p}$ such that $\sum\limits_{i=1}^{p} |S_{i}| \geq k$.
If the closed neighbourhood of $S_i$ does not intersect  $S_j$ 
 for all  $i\neq j$ then $G$ has a locally minimal defensive alliance of 
 size at least $k$.
\end{observation}
It is easy to see that $S = \bigcup\limits_{i=1}^{p} S_{i}$ is a locally minimal defensive alliance of size at least $k$.\\

\noindent Next, we introduce the concept of crucial vertices in a given defensive alliance which is useful to provide a characterization of locally minimal defensive alliances.

\begin{definition}\label{crucial}\rm
Let $D$ be a defensive alliance in $G$ with $\delta(G)\geq 2$. A vertex  $u\in D$ is said to be a \emph{crucial} vertex in $D$ if $u$ has a marginally protected neighbour $u_{1}$ in $D$ and $u_{1}$  also has a  marginally protected neighbour $u_{2}$ in $D$ ($u_{2}$ and $u$ are not necessarily distinct vertices). 
\end{definition} 

\begin{lemma}\label{Def}\rm  Let $G$ be a graph with $\delta(G)\geq 2$.
A defensive alliance $D$ in $G$ is a locally minimal defensive alliance if and only if all the vertices in $D$ are crucial.
\end{lemma}

\proof First we note that $|D|\geq 2$ as $\delta(G)\geq 2$. In the forward part, let us consider any vertex $u$ in the locally minimal defensive alliance $D$.
Since $D \setminus \{u\}$ is not a defensive alliance, it implies that $u$ must have a marginally protected neighbour in $D$.
Let us call it $u_{1}$. 
Due to the same reason $u_{1}$ must also have a marginally protected neighbour $u_{2}$ in $D$. It implies that  $u$ is a crucial vertex in $D$.
In the backward part, one can observe that $D\setminus \{v\}$ cannot form a defensive alliance for any $v\in D$ as every vertex $v$ is adjacent to a marginally protected vertex in $D$.  \qed\\

\par Now we formalize the idea in Observation \ref{obs1} in the form of an algorithm. Algorithm \ref{algo1} starts with a defensive alliance and returns a locally minimal defensive alliance. 

\begin{algorithm}[ht]
\caption{}\label{algo1}
\begin{algorithmic}[1]
\Require A graph $G=(V,E)$ and a defensive alliance $D$
\Ensure a locally minimal defensive alliance 
\For{$u\in D$}
\If{all neighbours of $u$ in $D$ are  overprotected} 
    \State $D=D\setminus \{u\}$
\EndIf
\EndFor

\State return $D$
\end{algorithmic}
\end{algorithm}

\begin{lemma}\rm \label{cruciallemma}
Let $D$ be a defensive alliance in graph $G$ and
let $S$ be the set of crucial vertices in $D$. Then the locally minimal 
defensive alliance obtained by Algorithm \ref{algo1} on the input instance  $(G,D)$ 
 always contains $S$.
\end{lemma}

\proof Let $u$ be a crucial vertex in $D$, that is, $u$ has a marginally protected neighbour $u_{1}$ in $D$ and $u_{1}$  also has a  marginally protected neighbour $u_{2}$ in $D$. First, we  argue that Algorithm \ref{algo1} cannot delete the vertices $u_{1}$ and $u_{2}$. For the sake of contradiction, let us assume that  $u_{1}$ was removed at some iteration during the execution of the algorithm. 
This implies that all the neighbours of $u_{1}$ in $D$ at that iteration were overprotected. Note that as vertex $u_{2}$ is a marginally protected neighbour of $u_{1}$ in $D$, it must have got deleted before $u_{1}$. Now, symmetrically we can argue that the vertex $u_{1}$ was deleted before $u_{2}$. This is a contradiction. Therefore Algorithm \ref{algo1} cannot delete $u_1$ or $u_2$. Again, for the sake of contradiction, let us assume that $u$ was removed at some iteration during the execution of the algorithm. This would imply that all the neighbours of $u$ in $D$ at that iteration were overprotected. Since $u_{1}$ is a marginally protected neighbour of $u$ in $D$, it must have got deleted before $u$. This is a contradiction as Algorithm \ref{algo1} cannot delete $u_{1}$. \qed 

\vspace{5pt}

\noindent Next we modify Algorithm \ref{algo1}. 
Note that in  Algorithm \ref{algo1}, the vertices from $D$ were getting deleted in an arbitrary order. In the following algorithm, we put some  restrictions on the order in which the vertices are deleted from $D$ to obtain a locally minimal defensive alliance.

\begin{algorithm}
\caption{\sc  }\label{alg:cap}
\begin{algorithmic}[1]
\Require A graph $G=(V,E)$, a defensive alliance $D\subseteq V(G)$ and a set $C\subseteq D$.
\Ensure a locally minimal defensive alliance.
\For{$u\in D\setminus C$}
\If{all neighbours of $u$ in $D$ are   overprotected} 
    \State $D=D\setminus \{u\}$
\EndIf
\EndFor
\State run Algorithm \ref{algo1} on $(G,D)$.
\end{algorithmic}
\end{algorithm}

\begin{lemma}\rm\label{cruciallemma2}
Let $D_{C}$ be the defensive alliance obtained at the end of {\bf for} loop in Algorithm \ref{alg:cap}. Then every vertex in  $D_{C} \setminus N[C]$ is a crucial vertex in the defensive alliance $D_{C}$. In other words, every vertex $v \in D_{C}$ such that $d(v,C) \geq 2$ is a crucial vertex in the defensive alliance $D_{C}$.
\end{lemma}

\proof Let $u$ be an arbitrary vertex in $D_{C} \setminus N[C]$.
Since every vertex in $D_{C} \setminus C$ has a marginally protected neighbour, 
$u$ has a marginally protected neighbour $u_1$ in $D_{C}$.
Clearly, $u_1$ is in  $D_{C}\setminus C$.
Therefore $u_1$ also  has a marginally protected neighbour $u_2$ 
($u_2$ can be $u$) in $D_{C}$.  By Definition \ref{crucial}, $u$ is crucial in $D_{C}$. 
Therefore every vertex in $D_{C} \setminus N[C]$  is a crucial vertex. This completes the proof of the lemma. \qed 

\begin{lemma}\rm \label{diameter}
Let $G$ be a simple undirected graph with minimum degree at least 2
and $k$ be a positive integer. 
If  $diam(G) \geq 8k$ then $G$ has a locally minimal defensive alliance of size at least $k$.
\end{lemma}

\proof As the $diam(G) \geq 8k$, there are two vertices $v_{0}$ and $v_{8k}$ such that $d(v_{0},v_{8k})=8k$. 
 Let $P=<v_{0},v_{1},\ldots,v_{8k}>$ be a shortest  $v_{0}-v_{8k}$ path in $G$. Set $C=\{v_{4i} ~|~ 0 \leq i \leq 2k\}$; note that the distance between
 any two vertices
 of $C$ is at least 4. 
Run Algorithm \ref{alg:cap} on $(G,V(G),C)$.
 Let $D_{C}$ be a defensive alliance obtained at the end of the {\bf for} loop 
 of lines 1-5 in Algorithm \ref{alg:cap}. Note that $D_{C}$ may not be a connected defensive alliance. Suppose $S_{1},S_{2},\ldots,S_{p}$ are the connected components of $D_C$. 
This implies that $D_{C}$ generates a partition of $C$ into $p$ parts $C_{1},C_{2},\ldots,C_{p}$ such that $C_{i} \subseteq S_{i}$. Line 6 of 
Algorithm \ref{alg:cap} executes 
Algorithm \ref{algo1} on $(G,D_C)$, that is, it executes Algorithm \ref{algo1} on $(G,S_i)$ for all $i$. 

\begin{claim}
Line 6 of Algorithm \ref{alg:cap} runs Algorithm \ref{algo1} on  $(G,S_{i})$ and
outputs a locally minimal defensive alliance $S'_{i}$ such that $|S'_{i}| \geq \max\big\{2,|C_{i}|-1\big\}$ for all $1\leq i \leq p$.
\end{claim}
Due to Observation \ref{obs1}, we always have $|S'_{i}|\geq 2$. 
We now prove that Algorithm \ref{algo1} on input instance $(G,S_i)$ returns a locally minimal
defensive alliance $S'_i$ of size at least $|C_i|-1$. Suppose $C_i=\{c_1, \ldots, c_q \}$. We know $d(c_j, c_{j+1})\geq 4$ and  $d(v_{0},c_{j}) <  d(v_{0},c_{j+1})$ for all $1\leq j \leq q-1$.  
As $G[S_i]$ is connected, 
a shortest path between $c_j$ and $c_{j+1}$ in $G[S_i]$ contain a vertex, say 
$x$,  such that $x$ is at distance at least 2 from both $c_j$ and $c_{j+1}$, that is, $d(x,C_i)\geq 2$. By Lemma \ref{cruciallemma2}, $x$ is a 
crucial vertex. As $C_i$ has $q$ vertices, we have at least $q-1$ crucial vertices in $S_i$. 
By Lemma \ref{cruciallemma}, the locally minimal defensive alliance $S_i'$ returned 
by line 6 of Algorithm \ref{alg:cap} contains all $q-1$ crucial vertices of $S_i$. 
Thus we have $|S'_{i}| \geq q-1$. This completes the proof of the claim.


\noindent As the closed neighbourhood of $S'_i$ does not intersect  $S'_j$, 
 for all  $i\neq j$, by Observation \ref{Union1} we get $\bigcup\limits_{i=1}^{p} {S'_i}$ is a locally minimal defensive alliance of size at least $k$. This completes the proof of the lemma. \qed

\vspace{5pt} 
\noindent Note that the above construction of locally minimal defensive alliance takes polynomial time as Algorithm \ref{algo1} and Algorithm \ref{alg:cap} runs in polynomial time. Also with more careful analysis in the proof of Lemma \ref{diameter}, the upper bound on the diameter of graph can be improved to $2k$. As this does not change the asymptotic complexity in the final algorithm, we present simple analysis.

 \subsection{Proof of Theorem \ref{thm:lmda}}
By Lemma \ref{diameter}, we know if  $diam(G) \geq 8k$, then $G$ has a  locally minimal defensive alliance of size at least $k$. So we assume $diam(G) < 8k$. 
The maximum degree of $G$ is $\Delta$. Therefore the number of vertices in $G$ is at most ${\Delta}^{\mathcal{O}(k)}$. This completes the proof of the theorem. \qed\\

\begin{corollary}\rm \label{cor:lmda}
Let $G$ be a simple  undirected  graph  with $\delta(G)\geq 2$ containing a connected defensive alliance $S$,  and $k$ be  a positive integer. If $diam(G[S])\geq 8k$, then $G$ has a locally minimal defensive alliance of size at least $k$.
\end{corollary}

\proof We use the same argument as before;  run Algorithm \ref{alg:cap} on
$(G,S,C)$ where $C \subseteq S$ instead of $(G,V(G),C)$. \qed

\subsection{Proof of Theorem \ref{thm:clmda}}

Our goal here is to find  a connected locally minimal defensive alliance $S$ of size at 
least $k$ in $G$. 
First we  guess a vertex $v$ of  $S$. There are at most $n$ candidates 
for $v$.
Therefore our goal now is to find a connected locally minimal defensive alliance containing vertex $v$, if it exists.
To do this, we  run  breadth first search on $G$ and get the BFS tree $T$ rooted at $v$.
Let us assume that the BFS tree $T$ has height $d$. Let $L_i$ be the set of
vertices at level $i$ of $T$ for $0\leq i\leq d-1$.
We guess the intersection of solution with the first $k+3$ levels of $T$, that is, level 0 to
level $k+2$. 
As the maximum degree is bounded by $\Delta$, the number of vertices in the first 
$k+3$ levels of the BFS tree is at most $\Delta^{k+2}$. Hence the
 number of guesses is bounded by $2^{\Delta^{k+2}}$.
Let us denote the intersection as $S'=S \cap \bigcup\limits_{i=0}^{k+2} L_{i}$. 
Note that $G[S']$ may not be connected. Let $S'_v$ be the connected component  containing $v$;
$S'_v$ must include at least $k$ vertices. A vertex  $x\in S'_v$ is said to be \emph{crucial} in $S'_v$ if $x$ has a marginally protected neighbour $x_1$ in $S'_v$ and $x_1$  also has a marginally protected neighbour $x_2$ in $S'_v$ ($x_2$ and $x$ are not necessarily distinct vertices). 
In polynomial time, we can check if each vertex of 
$S'\cap \bigcup\limits_{i=0}^{k-1} L_{i}$ is crucial or not. We say $S'$ is a valid guess if every vertex of $S'\cap \bigcup\limits_{i=0}^{k-1} L_{i}$ is crucial and protected 
in $S'_v$. To prove this theorem, we need the following simple claim.


\begin{claim}
 If  there exists a defensive alliance $D$ such that 
 $D \cap \bigcup\limits_{i=0}^{k+2} L_{i} = S'$ and $S'$ is a valid guess then there 
 exists a connected locally minimal defensive alliance $S$ of size at least $k$ 
 and it can be obtained in polynomial time. 
\end{claim}

\noindent{\it Proof of Claim:} Run Algorithm \ref{algo1} on $D$; it will return a locally minimal 
defensive alliance $S$. As $S'$ is a valid guess, the vertices in $S' \cap \bigcup\limits_{i=0}^{k-1} L_{i} $  are crucial and hence 
will not be removed by Algorithm \ref{algo1}. Thus $S$ is a connected locally minimal
defensive alliance of size at least $k$. This proves the claim.


\noindent Now we see how to find a defensive alliance $D$ such that 
 $D \cap \bigcup\limits_{i=0}^{k+2} L_{i} = S'$ and $S'$ is a valid guess.
 We begin with the set $D'' = S' \bigcup\limits_{i=k+3}^{d-1} L_{i}$.
 Note that $D''$ may not be a defensive alliance. We keep removing the unprotected vertices from $D''$ until all the vertices are protected.
Let $D'$ be the defensive alliance obtained by this method.
Note that $D'$ may not be a connected defensive alliance. 
Now, consider the connected defensive alliance $D$ obtained from $D'$ by taking the connected component in $G[D']$ containing $v$.
Note that the vertices of $C=D\cap \bigcup\limits_{i=0}^{k-1} L_{i}$
are crucial in $D$ and as otherwise we can 
confirm that our guess is wrong. 
Due to Lemma \ref{cruciallemma}, execution of Algorithm \ref{algo1} on $D$  generates a locally minimal defensive alliance containing $C$. Since $G[C]$ is connected, we will obtain a connected locally minimal defensive alliance containing $C$ and therefore containing $v$ as well. Since $|C|\geq k$, we obtain a connected locally minimal defensive alliance of size at least $k$, 
containing $v$.\qed\\

\section{FPT algorithm for {\sc Locally Minimal Defensive Alliance} parameterized by $k$}

\noindent In this section we  present an FPT algorithm for {\sc Locally Minimal Defensive Alliance} when the parameter is the solution size $k$. 
\begin{theorem}\label{f(k)} The 
{\sc Locally Minimal Defensive Alliance} problem restricted to  graphs of minimum 
degree at least 2,  admits
a kernel with at most ${f(k)}$ vertices for some computable function $f(k)$.
\end{theorem}

\proof  By Lemma \ref{diameter}, if  $diam(G) \geq 8k$, then $G$ has a  locally minimal defensive alliance of size at  least $k$. So we assume $diam(G) < 8k$. 
In order to prove this theorem,  we need the following claim:

\begin{claim}  Let $G$ be a graph with $diam(G) < 8k$. If $G$ has a vertex of degree greater than or equal to  a sufficiently large computable function $f_0(k)$ then $G$ has a locally minimal defensive alliance of size at least $k$.
\end{claim}

\noindent {\it Proof of Claim:} Suppose $d(u_0)=f_{0}(k)$ for some sufficiently large computable function $f_{0}(k)$. Run Algorithm \ref{alg:cap} on $(G,V(G), \{u_0\})$ and suppose it returns $D_0$.\\
Case 1: If $u_0$ is in $D_0$ then at least $\frac{d(u_0)}{2}>k$ neighbours  of $u_0$ are in $D_0$ for its protection. That means, we have a locally minimal defensive alliance $D_0$ of size at 
least $k$; so we are done. \\
Case 2: Suppose $u_0$ is not in $D_0$.  
Let $D'_0$ be the defensive alliance obtained at the end of the {\bf for} loop of 
lines 1-5 in Algorithm \ref{alg:cap} and 
all neighbours of $u_0$ in $D'_0$ are overprotected. 
This is why $u_0$ is deleted from $D'_0$. 
We partition the vertices of defensive alliance $D'_0\setminus \{u_{0}\}$ into two parts:
part $P_1^0$ contains vertices adjacent to $u_0$ and  part $P_2^0$ contains vertices not adjacent to $u_0$. Observe that the vertices of $P_2^0$  will be there in final locally minimal defensive alliance as they are crucial in $D'_0$.
Thus we have $|P_{2}^0|\leq k-1$, otherwise we have a solution of size at least $k$.
Furthermore, every vertex in $P_2^0$ has degree at most $2k$ or else we have a solution of size  at least $k$.
As $f_0(k)$ is a very large function and every vertex in $P_2^0$ has degree at most $2k$, 
we prove that $P_1^0$ contains at least one vertex $u_1$ with degree at least $f_{1}(k)$ for some large function $f_1$. Note that 
$G[D'_0\setminus \{u_0\}]$  contains at most $\frac{k}{2}-1$ connected components of size at
least two, otherwise by Observation \ref{obs1} and \ref{Union1}, it is a yes-instance.
Moreover,  every component of 
$G[D'_0\setminus \{u_0\}]$ has diameter at most $8k$, otherwise by Corollary \ref{cor:lmda},
it is a yes instance. Therefore, if the maximum degree of $G[D'_0\setminus \{u_0\}]$ is $f_{1}(k)$ then we know that it contains at most $\mathcal{O}(k.f_{1}(k)^{k})$ vertices. We also know that $G[D'_0\setminus \{u_0\}]$ contains at least $\mathcal{O}(f_{0}(k))$ vertices. Therefore, we get $c_1 f_{\mathcal{0}(k)}\leq c_2 k.f_1(k)^{k}$ where $c_1$ and $c_2$ are two positive real numbers. This implies 
$$f_1(k) \geq \frac{c_1}{c_2} \Big(\frac{f_0(k)}{k}\Big)^\frac{1}{k}. $$
Since we have control over the function $f_{0}(k)$, we can assume that $f_{1}(k)$ is as  large as we want by taking $f_{0}(k)$ large enough. 
\par We again run Algorithm \ref{alg:cap} on $(G,D'_0\setminus \{ u_0\}, \{u_1\})$. We can apply the same argument as before and either obtain a locally minimal defensive alliance of size at least $k$ containing $u_{1}$ or a defensive alliance $D'_{1} \setminus \{u_1\}$ containing a vertex $u_{2}$ of degree $f_{2}(k)$ for some large function $f_2(k)$.
\par We repeat the same procedure $k$ times. Let  $u_{k}$ be the vertex obtained 
by the same  procedure with degree at least $f_{k}(k)$ for some large computable  function $f_{k}$. Call Algorithm \ref{alg:cap} on $(G,D'_{k-1}\setminus \{u_{k-1}\},u_{k})$
and suppose it returns $D_{k}$.
If $u_k$ is in $D_k$ then at least $\frac{d(u_k)}{2}>k$ neighbours  of $u_k$ are in $D_k$ for its protection. That means, we have a locally minimal defensive alliance $D_k$ of size at 
least $k$; so we are done. Suppose $u_k$ is not in $D_k$. 
Let $D'_{k}$ be the defensive alliance obtained at the end of the {\bf for} loop of Algorithm \ref{alg:cap}.
We assume that all neighbours of  $u_{k}$ in $D'_{k}$ are overprotected and this is why
$u_{k}$ is deleted.  
We partition the vertices of defensive alliance $D'_{k}\setminus \{u_{k}\}$ into two parts: 
$P^{k}_1$ and $P^{k}_2$.
A vertex $x$ is in part $P^{k}_{1}$ if it is adjacent to all $u_i$'s  and 
a vertex $x$ is in part $P^{k}_{2}$ if it is not adjacent to at least one $u_i$ for some $0 \leq i \leq k$.
By Lemma \ref{cruciallemma2}, the vertices of $P^{k}_2$ are crucial and by Lemma \ref{cruciallemma}, these crucial vertices will be there in the
final locally minimal defensive alliance. 
If $|P^{k}_2|\geq k$, then we have a locally minimal defensive
alliance of size at least $k$ and we are done. Thus we assume $|P^{k}_2|<k$.
Consider a vertex  $x \in P^{k}_{1}$. By the definition of $ P^{k}_{1}$,
$x$ is adjacent to $k+1$ vertices $u_0,u_1,\ldots,u_{k}$ which are 
outside the defensive alliance $D'_{k}\setminus \{u_{k}\}$. As $x$ is protected in
$D'_{k}\setminus \{u_{k}\}$ it must have at least $k$ neighbours in $D'_{k}\setminus \{u_{k}\}$. Thus $d_{G}(x)\geq 2k+1$. 
Line 6 of Algorithm \ref{alg:cap} executes  Algorithm \ref{algo1} on $(G, D'_{k}\setminus \{u_{k}\})$; suppose it outputs $S$. We prove  that at least one vertex of $P^{k}_{1}$ will survive in $S$.
For the sake of contradiction, assume that this is false, that is, no vertex
survive in $S$. Suppose vertex $x$ in $P^{k}_{1}$ is  deleted last. It has at most $k$ neighbours, including itself, in 
the current defensive alliance as $P^{k}_{2}$  contains at most $k-1$ vertices and $k+1$ neighbours outside the current defensive alliance. That means, $x$ is not protected, a contradiction to the fact that $x$ is protected in $D'_{k}\setminus \{u_{k}\}$. Therefore, the locally minimal defensive alliance $S$ obtained by Algorithm 
\ref{alg:cap} contains a vertex of degree at least $2k+1$; hence $S$ is of size at
least $k$.  This completes the proof of the claim. \\

\noindent By the above claim, we know if $\Delta(G)\geq f_0(k)$ where $f_0(k)$ is a large computable function, then we have a locally minimal defensive alliance of size at least $k$. Thus we can assume $\Delta(G)\leq f_0(k)$. As $G$ has diameter at most $8k$ and 
 $\Delta(G)\leq f_0(k)$, $G$ admits a kernel of size at most $f_0(k)^{8k}$. This completes the proof of Theorem 
 \ref{f(k)}. \qed\\\\
\noindent For the sake of completeness, we note that having $f_0(k)$ at least  $k^{{k}^{\mathcal{O}(k)}}$ is enough to make the argument work in the above proof. In the following section, we improve this function $f_0(k)$ when restricted to some special graph classes. 

\section{Kernels for {\sc Locally Minimal Defensive Alliance} restricted to $C_3$-free and $C_4$-free graphs}\label{special kernels}

\noindent In this section we give improved kernels for {\sc Locally Minimal Defensive Alliance} restricted to $C_3$-free and $C_4$-free graphs, when the parameter is the solution size $k$. The following lemma is the basis for our kernelization algorithm. 
\begin{lemma}\rm \label{C3}
Let $G$ be a $C_3$-free  graph with minimum degree at least 2. If $\Delta(G) \geq 4k^{2}$ then $G$ has a locally minimal defensive alliance of size at least $k$.
\end{lemma}
\proof
 Suppose $d(u)= 4k^2$. Run BFS$(G,u)$ to obtain the BFS tree
$T$ rooted at $u$ and to get level for each vertex in $G$. Let $L_i$ be the set of
vertices at level $i$ of $T$. Run Algorithm \ref{alg:cap} on $(G,V(G),u)$ and suppose it outputs $D$. If $u \in D$ then clearly $D$ is a locally minimal defensive alliance of size at least $2k^2$ as degree of $u$ is $4k^2$ and $u$ is protected in $D$. 
Consider the case where $u$ is not in $D$.  
Let $D'$ be the defensive alliance obtained at line 5 of Algorithm \ref{alg:cap}.
 As $u$ is protected in $D'$ at least $2k^{2}$ of its neighbours are in 
$L_1\cap D'$.  All neighbours of $u$ in $L_1\cap D'$ are overprotected, this is why $u$ is deleted from $D'$ in line 6. 
As $G$ is triangle-free and  $\delta(G) \geq 2$, every  $x\in L_1\cap D'$ has least one neighbour in $L_2 \cap D'$.
We know  every vertex from level $2$ or higher that survives in the defensive alliance obtained by Algorithm \ref{alg:cap} are crucial and they remain  in the final locally minimal defensive alliance.
This implies that there are at most $k-1$ vertices in 
$ D' \cap L_2$, otherwise we have a solution of size at least $k$.
 Note that there are at least $2k^2$ vertices in $D'\cap L_1$, at most $k-1$ vertices in $L_2\cap D'$ and  each vertex in $D'\cap L_1$ has at least one neighbour in  $L_2\cap D'$. 
 By Pigeonhole principle there exists a vertex $w \in L_2\cap D'$ of degree at least $2k$.
As $w$ is crucial it is in the final solution. Moreover, as $d(w)\geq 2k$, the final locally minimal defensive alliance is
of size at least $k$.

\begin{lemma}\rm \label{C4}
Let $G$ be a $C_4$-free graph with minimum degree at least 2. If $\Delta(G) \geq 4k^{2}-2k$ then $G$ has a locally minimal defensive alliance of size at least $k$.
\end{lemma}
\proof
 Let $d(u)\geq 4k^2-2k$. Run BFS$(G,u)$ to obtain the BFS tree
$T$ rooted at $u$ and to get level for each vertex in $G$. Run Algorithm \ref{alg:cap} on $(G,V(G),u)$ and suppose it outputs $D$. If $u \in D$ then clearly $D$ is a locally minimal defensive alliance of size at least $2k^2-k$ as degree of $u$ is $4k^{2}-2k$ and $u$ is protected in $D$. 
Consider the case where $u$ is not in $D$.  
Let $D'$ be the defensive alliance obtained at line 5 of Algorithm \ref{alg:cap}.
As $u$ is protected in $D'$ at least $2k^{2}-k$ of its neighbours are in 
$L_1\cap D'$.  All neighbours of $u$ in $L_1\cap D'$ are overprotected, this is why $u$ is deleted from $D'$ in line 6. 
Now let us focus on $G[L_1 \cap D']$. It may be noted that 
the maximum degree of a vertex in $G[L_1 \cap D']$ is at most 1 as  $G$ is $C_{4}$-free. 
As $\delta(G)\geq 2$, all isolated vertices in $G[L_1 \cap D']$ must have at least one neighbour in $L_{2} \cap D'$. Every vertex from level $2$ or higher that survives in the defensive alliance obtained by Algorithm \ref{alg:cap} are crucial and they remain  in the final locally minimal defensive alliance.
Thus we assume there are at most $k-1$ vertices in 
$ D' \cap L_2$, otherwise we have a solution of size at least $k$.
Furthermore, every vertex in $ D' \cap L_2$ has degree at most $2k$ or else we have a solution of size  at least $k$.
This  implies that there are at most $2k^{2}-2k$ vertices in $G[N(u) \cap D']$ which have neighbour in $L_{2} \cap D'$.
Note that there are still $k$ vertices which have no neighbour in $L_{2} \cap D'$; so they must have at least one neighbour in $L_{1} \cap D'$. 
As $G[N(u) \cap D']$ has maximum degree one, we must have at least $\frac{k}{2}$ isolated edges in $G[D'\setminus \{u\}]$.
As  these $\frac{k}{2}$ isolated edges in $G[D'\setminus \{u\}]$ are connected defensive alliances such that closed neighbour of one defensive alliance does not intersect other, by Observation \ref{Union1}   we get a locally minimal defensive alliance of size at least $k$. \qed\\\\

\begin{theorem}\label{C3C4free} The {\sc Locally Minimal Defensive Alliance} problem 
parameterized by $k$ restricted to  $C_3$-free or $C_4$-free graphs 
of minimum degree at least 2, admits
a kernel with at most $k^{\mathcal{O}(k)}$ vertices.
\end{theorem}

\proof If $\Delta(G)\geq 4k^{2}$ or $diam(G)\geq 8k$ then we get a yes instance due to Lemma \ref{C3}, Lemma \ref{C4}  and Lemma \ref{diameter}. Therefore, we assume $G$ has $\Delta(G)< 4k^{2}$ and $diam(G)< 8k$ which produces a kernel of size $k^{\mathcal{O}(k)}$.\qed

\section{FPT algorithm for {\sc Locally Minimal Defensive Alliance} on planar graphs}\label{planar FPT}

It is proved in  \cite{DBPL-LMDA} that the {\sc Locally Minimal Defensive Alliance} is NP-complete in planar graphs, via a reduction from {\sc Minimum Maximal Matching}  in cubic planar graph. In this section, we design an FPT algorithm for {\sc Locally Minimal Defensive Alliance} on planar graphs. We use win/win approach to design an FPT algorithm for 
 {\sc Locally Minimal Defensive Alliance} on planar graphs. 
 For $n \in \mathbb{N}$, by $[n]$ we denote the set $\{1,2,\ldots,n\}$.
 For a positive integer $t$,
 a $t\times t$ \emph{grid} $\boxplus_t$ is a graph with vertex set $\{(x,y)~:~x,y\in [t]\}$ and two 
 different vertices $(x,y)$ and $(x',y')$ are adjacent if and only if 
 $|x-x'|+|y-y'|=1$. The triangulated grid $\Gamma_t$ is obtained from the grid $\boxplus_t$
by adding the edges $(x+1,y),(x,y+1)$ for all $1\leq x,y\leq t-1$
and additionally making vertex $(t,t)$ adjacent to the whole border of $\boxplus_t$.
Theorem \ref{planargrid} gives the relationship between the treewidth and and the size of a
tringulated grid as a contraction.
\begin{theorem} \label{planargrid}\cite{FOMIN2011302}
{\bf(Planar excluded grid theorem for edge contractions)}.
For every connected planar graph $G$ and integer $t > 0$, if $tw(G) > 9t+5$ then $G$ contains $\Gamma_{t}$ as a contraction.
Furthermore, for every $\epsilon >0$ there exists an $\mathcal{O}(n^{2})$ algorithm that, given a connected planar $n$-vertex graph $G$
and integer $t$, either outputs a tree decomposition of $G$ of width $(9 + \epsilon)t + 5$ or a set of edges whose contraction in G results in $\Gamma_{t}$.
\end{theorem}

\begin{figure}[ht]
    \centering
    \begin{tikzpicture}[scale=0.5]
    \node[circle, draw, inner sep=0pt,fill=black, minimum size=4pt] (u00) at (0,-.8) [label=left:] {};
\node[circle, draw, inner sep=0pt,fill=black, minimum size=4pt] (u01) at (.8,-.8) [label=left:] {};
\node[circle, draw, inner sep=0pt,fill=black, minimum size=4pt] (u02) at (1.6,-.8) [label=left:] {};
\node[circle, draw, inner sep=0pt,fill=black, minimum size=4pt] (u03) at (2.4,-.8) [label=left:] {};
\node[circle, draw, inner sep=0pt,fill=black, minimum size=4pt] (u04) at (3.2,-.8) [label=left:] {};
\node[circle, draw, inner sep=0pt,fill=black, minimum size=4pt] (u05) at (4,-.8) [label=left:] {};
\node[circle, draw, inner sep=0pt,fill=black, minimum size=4pt] (u06) at (4.8,-.8) [label=left:] {};
\node[circle, draw, inner sep=0pt,fill=black, minimum size=4pt] (u07) at (5.6,-.8) [label=left:] {};
\node[circle, draw, inner sep=0pt,fill=black, minimum size=4pt] (u08) at (6.4,-.8) [label=left:] {};
\node[circle, draw, inner sep=0pt,fill=black, minimum size=4pt] (u09) at (7.2,-.8) [label=left:] {};
\node[circle, draw, inner sep=0pt,fill=black, minimum size=4pt] (u010) at (8,-.8) [label=left:] {};
\node[circle, draw, inner sep=0pt,fill=black, minimum size=4pt] (u011) at (8.8,-.8) [label=left:] {};
\node[circle, draw, inner sep=0pt,fill=black, minimum size=4pt] (u012) at (9.6,-.8) [label=left:] {};
\node[circle, draw, inner sep=0pt,fill=black, minimum size=4pt] (u013) at (10.4,-.8) [label=left:] {};

\node[circle, draw, inner sep=0pt,fill=black, minimum size=4pt] (u10) at (0,0) [label=left:] {};
\node[circle, draw, inner sep=0pt,fill=black, minimum size=4pt] (u11) at (.8,0) [label=left:] {};
\node[circle, draw, inner sep=0pt,fill=black, minimum size=4pt] (u12) at (1.6,0) [label=left:] {};
\node[circle, draw, inner sep=0pt,fill=black, minimum size=4pt] (u13) at (2.4,0) [label=left:] {};
\node[circle, draw, inner sep=0pt,fill=black, minimum size=4pt] (u14) at (3.2,0) [label=left:] {};
\node[circle, draw, inner sep=0pt,fill=black, minimum size=4pt] (u15) at (4,0) [label=left:] {};
\node[circle, draw, inner sep=0pt,fill=black, minimum size=4pt] (u16) at (4.8,0) [label=left:] {};
\node[circle, draw, inner sep=0pt,fill=black, minimum size=4pt] (u17) at (5.6,0) [label=left:] {};
\node[circle, draw, inner sep=0pt,fill=black, minimum size=4pt] (u18) at (6.4,0) [label=left:] {};
\node[circle, draw, inner sep=0pt,fill=black, minimum size=4pt] (u19) at (7.2,0) [label=left:] {};
\node[circle, draw, inner sep=0pt,fill=black, minimum size=4pt] (u110) at (8,0) [label=left:] {};
\node[circle, draw, inner sep=0pt,fill=black, minimum size=4pt] (u111) at (8.8,0) [label=left:] {};
\node[circle, draw, inner sep=0pt,fill=black, minimum size=4pt] (u112) at (9.6,0) [label=left:] {};
\node[circle, draw, inner sep=0pt,fill=black, minimum size=4pt] (v113) at (10.4,0) [label=left:] {};

\node[circle, draw, inner sep=0pt,fill=black, minimum size=4pt] (u20) at (0,.8) [label=left:] {};
\node[circle, draw, inner sep=0pt,fill=black, minimum size=4pt] (u21) at (.8,.8) [label=left:] {};
\node[circle, draw, inner sep=0pt,color=red,fill=red, minimum size=4pt] (u22) at (1.6,.8) [label=left:] {};
\node[circle, draw, inner sep=0pt, fill=black,minimum size=4pt] (u23) at (2.4,.8) [label=left:] {};
\node[circle, draw, inner sep=0pt, fill=black,minimum size=4pt] (u24) at (3.2,.8) [label=left:] {};
\node[circle, draw, inner sep=0pt, fill=black,minimum size=4pt] (u25) at (4,.8) [label=left:] {};
\node[circle, draw, inner sep=0pt,fill=black, minimum size=4pt] (u26) at (4.8,.8) [label=left:] {};
\node[circle, draw, inner sep=0pt,fill=black, minimum size=4pt] (u27) at (5.6,.8) [label=left:] {};
\node[circle, draw,fill=black, inner sep=0pt, minimum size=4pt] (u28) at (6.4,.8) [label=left:] {};
\node[circle, draw,fill=black, inner sep=0pt, minimum size=4pt] (u29) at (7.2,.8) [label=left:] {};
\node[circle, draw, fill=black,inner sep=0pt, minimum size=4pt] (u210) at (8,.8) [label=left:] {};
\node[circle, draw, fill=black,inner sep=0pt, minimum size=4pt] (u211) at (8.8,.8) [label=left:] {};
\node[circle, draw, fill=black, inner sep=0pt, minimum size=4pt] (u212) at (9.6,.8) [label=left:] {};
\node[circle, draw, fill=black, inner sep=0pt, minimum size=4pt] (u213) at (10.4,.8) [label=left:] {};

\node[circle, draw, inner sep=0pt,fill=black, minimum size=4pt] (u30) at (0,1.6) [label=left:] {};
\node[circle, draw, inner sep=0pt,fill=black, minimum size=4pt] (u31) at (.8,1.6) [label=left:] {};
\node[circle, draw, inner sep=0pt,fill=black, minimum size=4pt] (u32) at (1.6,1.6) [label=left:] {};
\node[circle, draw, inner sep=0pt, fill=black,minimum size=4pt] (u33) at (2.4,1.6) [label=left:] {};
\node[circle, draw, inner sep=0pt, fill=black,minimum size=4pt] (u34) at (3.2,1.6) [label=left:] {};
\node[circle, draw, inner sep=0pt, fill=black,minimum size=4pt] (u35) at (4,1.6) [label=left:] {};
\node[circle, draw, inner sep=0pt,fill=black, minimum size=4pt] (u36) at (4.8,1.6) [label=left:] {};
\node[circle, draw, inner sep=0pt,fill=black, minimum size=4pt] (u37) at (5.6,1.6) [label=left:] {};
\node[circle, draw,fill=black, inner sep=0pt, minimum size=4pt] (u38) at (6.4,1.6) [label=left:] {};
\node[circle, draw,fill=black, inner sep=0pt, minimum size=4pt] (u39) at (7.2,1.6) [label=left:] {};
\node[circle, draw, fill=black,inner sep=0pt, minimum size=4pt] (u310) at (8,1.6) [label=left:] {};
\node[circle, draw, fill=black,inner sep=0pt, minimum size=4pt] (u311) at (8.8,1.6) [label=left:] {};
\node[circle, draw, fill=black, inner sep=0pt, minimum size=4pt] (u312) at (9.6,1.6) [label=left:] {};
\node[circle, draw, fill=black, inner sep=0pt, minimum size=4pt] (u313) at (10.4,1.6) [label=left:] {};

\node[circle, draw, inner sep=0pt,fill=black, minimum size=4pt] (u40) at (0,2.4) [label=left:] {};
\node[circle, draw, inner sep=0pt,fill=black, minimum size=4pt] (u41) at (.8,2.4) [label=left:] {};
\node[circle, draw, inner sep=0pt,fill=black, minimum size=4pt] (u42) at (1.6,2.4) [label=left:] {};
\node[circle, draw, inner sep=0pt, fill=black,minimum size=4pt] (u43) at (2.4,2.4) [label=left:] {};
\node[circle, draw, inner sep=0pt, fill=black,minimum size=4pt] (u44) at (3.2,2.4) [label=left:] {};
\node[circle, draw, inner sep=0pt, fill=black,minimum size=4pt] (u45) at (4,2.4) [label=left:] {};
\node[circle, draw, inner sep=0pt,fill=black, minimum size=4pt] (u46) at (4.8,2.4) [label=left:] {};
\node[circle, draw, inner sep=0pt,fill=black, minimum size=4pt] (u47) at (5.6,2.4) [label=left:] {};
\node[circle, draw,fill=black, inner sep=0pt, minimum size=4pt] (u48) at (6.4,2.4) [label=left:] {};
\node[circle, draw,fill=black, inner sep=0pt, minimum size=4pt] (u49) at (7.2,2.4) [label=left:] {};
\node[circle, draw, fill=black,inner sep=0pt, minimum size=4pt] (u410) at (8,2.4) [label=left:] {};
\node[circle, draw, fill=black,inner sep=0pt, minimum size=4pt] (u411) at (8.8,2.4) [label=left:] {};
\node[circle, draw, fill=black, inner sep=0pt, minimum size=4pt] (u412) at (9.6,2.4) [label=left:] {};
\node[circle, draw, fill=black, inner sep=0pt, minimum size=4pt] (u413) at (10.4,2.4) [label=left:] {};

\node[circle, draw, inner sep=0pt,fill=black, minimum size=4pt] (u50) at (0,3.2) [label=left:] {};
\node[circle, draw, inner sep=0pt,fill=black, minimum size=4pt] (u51) at (.8,3.2) [label=left:] {};
\node[circle, draw, inner sep=0pt,fill=black, minimum size=4pt] (u52) at (1.6,3.2) [label=left:] {};
\node[circle, draw, inner sep=0pt, fill=black,minimum size=4pt] (u53) at (2.4,3.2) [label=left:] {};
\node[circle, draw, inner sep=0pt, fill=black,minimum size=4pt] (u54) at (3.2,3.2) [label=left:] {};
\node[circle, draw, inner sep=0pt, fill=black,minimum size=4pt] (u55) at (4,3.2) [label=left:] {};
\node[circle, draw, inner sep=0pt,fill=black, minimum size=4pt] (u56) at (4.8,3.2) [label=left:] {};
\node[circle, draw, inner sep=0pt,fill=black, minimum size=4pt] (u57) at (5.6,3.2) [label=left:] {};
\node[circle, draw,fill=black, inner sep=0pt, minimum size=4pt] (u58) at (6.4,3.2) [label=left:] {};
\node[circle, draw,fill=black, inner sep=0pt, minimum size=4pt] (u59) at (7.2,3.2) [label=left:] {};
\node[circle, draw, fill=black,inner sep=0pt, minimum size=4pt] (u510) at (8,3.2) [label=left:] {};
\node[circle, draw, fill=black,inner sep=0pt, minimum size=4pt] (u511) at (8.8,3.2) [label=left:] {};
\node[circle, draw, fill=black, inner sep=0pt, minimum size=4pt] (u512) at (9.6,3.2) [label=left:] {};
\node[circle, draw, fill=black, inner sep=0pt, minimum size=4pt] (u513) at (10.4,3.2) [label=left:] {};

\node[circle, draw, inner sep=0pt,fill=black, minimum size=4pt] (u60) at (0,4) [label=left:] {};
\node[circle, draw, inner sep=0pt,fill=black, minimum size=4pt] (u61) at (.8,4) [label=left:] {};
\node[circle, draw, inner sep=0pt,fill=black, minimum size=4pt] (u62) at (1.6,4) [label=left:] {};
\node[circle, draw, inner sep=0pt, fill=black,minimum size=4pt] (u63) at (2.4,4) [label=left:] {};
\node[circle, draw, inner sep=0pt, fill=black,minimum size=4pt] (u64) at (3.2,4) [label=left:] {};
\node[circle, draw, inner sep=0pt, fill=black,minimum size=4pt] (u65) at (4,4) [label=left:] {};
\node[circle, draw, inner sep=0pt,fill=black, minimum size=4pt] (u66) at (4.8,4) [label=left:] {};
\node[circle, draw, inner sep=0pt,fill=black, minimum size=4pt] (u67) at (5.6,4) [label=left:] {};
\node[circle, draw,fill=black, inner sep=0pt, minimum size=4pt] (u68) at (6.4,4) [label=left:] {};
\node[circle, draw,fill=black, inner sep=0pt, minimum size=4pt] (u69) at (7.2,4) [label=left:] {};
\node[circle, draw, fill=black,inner sep=0pt, minimum size=4pt] (u610) at (8,4) [label=left:] {};
\node[circle, draw, fill=black,inner sep=0pt, minimum size=4pt] (u611) at (8.8,4) [label=left:] {};
\node[circle, draw, fill=black, inner sep=0pt, minimum size=4pt] (u612) at (9.6,4) [label=left:] {};
\node[circle, draw, fill=black, inner sep=0pt, minimum size=4pt] (u613) at (10.4,4) [label=left:] {};

\node[circle, draw, inner sep=0pt,fill=black, minimum size=4pt] (u70) at (0,4.8) [label=left:] {};
\node[circle, draw, inner sep=0pt,fill=black, minimum size=4pt] (u71) at (.8,4.8) [label=left:] {};
\node[circle, draw, inner sep=0pt,fill=black, minimum size=4pt] (u72) at (1.6,4.8) [label=left:] {};
\node[circle, draw, inner sep=0pt, fill=black,minimum size=4pt] (u73) at (2.4,4.8) [label=left:] {};
\node[circle, draw, inner sep=0pt, fill=black,minimum size=4pt] (u74) at (3.2,4.8) [label=left:] {};
\node[circle, draw, inner sep=0pt, fill=black,minimum size=4pt] (u75) at (4,4.8) [label=left:] {};
\node[circle, draw, inner sep=0pt,fill=black, minimum size=4pt] (u76) at (4.8,4.8) [label=left:] {};
\node[circle, draw, inner sep=0pt,fill=black, minimum size=4pt] (u77) at (5.6,4.8) [label=left:] {};
\node[circle, draw,fill=black, inner sep=0pt, minimum size=4pt] (u78) at (6.4,4.8) [label=left:] {};
\node[circle, draw,fill=black, inner sep=0pt, minimum size=4pt] (u79) at (7.2,4.8) [label=left:] {};
\node[circle, draw, fill=black,inner sep=0pt, minimum size=4pt] (u710) at (8,4.8) [label=left:] {};
\node[circle, draw, fill=black,inner sep=0pt, minimum size=4pt] (u711) at (8.8,4.8) [label=left:] {};
\node[circle, draw, fill=black, inner sep=0pt, minimum size=4pt] (u712) at (9.6,4.8) [label=left:] {};
\node[circle, draw, fill=black, inner sep=0pt, minimum size=4pt] (u713) at (10.4,4.8) [label=left:] {};

\node[circle, draw, inner sep=0pt,fill=black, minimum size=4pt] (u80) at (0,5.6) [label=left:] {};
\node[circle, draw, inner sep=0pt,fill=black, minimum size=4pt] (u81) at (.8,5.6) [label=left:] {};
\node[circle, draw, inner sep=0pt,fill=black, minimum size=4pt] (u82) at (1.6,5.6) [label=left:] {};
\node[circle, draw, inner sep=0pt, fill=black,minimum size=4pt] (u83) at (2.4,5.6) [label=left:] {};
\node[circle, draw, inner sep=0pt, fill=black,minimum size=4pt] (u84) at (3.2,5.6) [label=left:] {};
\node[circle, draw, inner sep=0pt, fill=black,minimum size=4pt] (u85) at (4,5.6) [label=left:] {};
\node[circle, draw, inner sep=0pt,fill=black, minimum size=4pt] (u86) at (4.8,5.6) [label=left:] {};
\node[circle, draw, inner sep=0pt,fill=black, minimum size=4pt] (u87) at (5.6,5.6) [label=left:] {};
\node[circle, draw,fill=black, inner sep=0pt, minimum size=4pt] (u88) at (6.4,5.6) [label=left:] {};
\node[circle, draw,fill=black, inner sep=0pt, minimum size=4pt] (u89) at (7.2,5.6) [label=left:] {};
\node[circle, draw, fill=black,inner sep=0pt, minimum size=4pt] (u810) at (8,5.6) [label=left:] {};
\node[circle, draw, fill=black,inner sep=0pt, minimum size=4pt] (u811) at (8.8,5.6) [label=left:] {};
\node[circle, draw, fill=black, inner sep=0pt, minimum size=4pt] (u812) at (9.6,5.6) [label=left:] {};
\node[circle, draw, fill=black, inner sep=0pt, minimum size=4pt] (u813) at (10.4,5.6) [label=left:] {};

\node[circle, draw, inner sep=0pt,fill=black, minimum size=4pt] (u90) at (0,6.4) [label=left:] {};
\node[circle, draw, inner sep=0pt,fill=black, minimum size=4pt] (u91) at (.8,6.4) [label=left:] {};
\node[circle, draw, inner sep=0pt,fill=black, minimum size=4pt] (u92) at (1.6,6.4) [label=left:] {};
\node[circle, draw, inner sep=0pt, fill=black,minimum size=4pt] (u93) at (2.4,6.4) [label=left:] {};
\node[circle, draw, inner sep=0pt, fill=black,minimum size=4pt] (u94) at (3.2,6.4) [label=left:] {};
\node[circle, draw, inner sep=0pt, fill=black,minimum size=4pt] (u95) at (4,6.4) [label=left:] {};
\node[circle, draw, inner sep=0pt,fill=black, minimum size=4pt] (u96) at (4.8,6.4) [label=left:] {};
\node[circle, draw, inner sep=0pt,fill=black, minimum size=4pt] (u97) at (5.6,6.4) [label=left:] {};
\node[circle, draw,fill=black, inner sep=0pt, minimum size=4pt] (u98) at (6.4,6.4) [label=left:] {};
\node[circle, draw,fill=black, inner sep=0pt, minimum size=4pt] (u99) at (7.2,6.4) [label=left:] {};
\node[circle, draw, fill=black,inner sep=0pt, minimum size=4pt] (u910) at (8,6.4) [label=left:] {};
\node[circle, draw, fill=black,inner sep=0pt, minimum size=4pt] (u911) at (8.8,6.4) [label=left:] {};
\node[circle, draw, fill=black, inner sep=0pt, minimum size=4pt] (u912) at (9.6,6.4) [label=left:] {};
\node[circle, draw, fill=black, inner sep=0pt, minimum size=4pt] (u913) at (10.4,6.4) [label=left:] {};

\node[circle, draw, inner sep=0pt,fill=black, minimum size=4pt] (u100) at (0,7.2) [label=left:] {};
\node[circle, draw, inner sep=0pt,fill=black, minimum size=4pt] (u101) at (.8,7.2) [label=left:] {};
\node[circle, draw, inner sep=0pt,fill=black, minimum size=4pt] (u102) at (1.6,7.2) [label=left:] {};
\node[circle, draw, inner sep=0pt, fill=black,minimum size=4pt] (u103) at (2.4,7.2) [label=left:] {};
\node[circle, draw, inner sep=0pt, fill=black,minimum size=4pt] (u104) at (3.2,7.2) [label=left:] {};
\node[circle, draw, inner sep=0pt, fill=black,minimum size=4pt] (u105) at (4,7.2) [label=left:] {};
\node[circle, draw, inner sep=0pt,fill=black, minimum size=4pt] (u106) at (4.8,7.2) [label=left:] {};
\node[circle, draw, inner sep=0pt,fill=black, minimum size=4pt] (u107) at (5.6,7.2) [label=left:] {};
\node[circle, draw,fill=black, inner sep=0pt, minimum size=4pt] (u108) at (6.4,7.2) [label=left:] {};
\node[circle, draw,fill=black, inner sep=0pt, minimum size=4pt] (u109) at (7.2,7.2) [label=left:] {};
\node[circle, draw, fill=black,inner sep=0pt, minimum size=4pt] (u1010) at (8,7.2) [label=left:] {};
\node[circle, draw, fill=black,inner sep=0pt, minimum size=4pt] (u1011) at (8.8,7.2) [label=left:] {};
\node[circle, draw, fill=black, inner sep=0pt, minimum size=4pt] (u1012) at (9.6,7.2) [label=left:] {};
\node[circle, draw, fill=black, inner sep=0pt, minimum size=4pt] (u1013) at (10.4,7.2) [label=left:] {};

\node[circle, draw, inner sep=0pt,fill=black, minimum size=4pt] (u110) at (0,8) [label=left:] {};
\node[circle, draw, inner sep=0pt,fill=black, minimum size=4pt] (u111) at (.8,8) [label=left:] {};
\node[circle, draw, inner sep=0pt,fill=black, minimum size=4pt] (u112) at (1.6,8) [label=left:] {};
\node[circle, draw, inner sep=0pt, fill=black,minimum size=4pt] (u113) at (2.4,8) [label=left:] {};
\node[circle, draw, inner sep=0pt, fill=black,minimum size=4pt] (u114) at (3.2,8) [label=left:] {};
\node[circle, draw, inner sep=0pt, fill=black,minimum size=4pt] (u115) at (4,8) [label=left:] {};
\node[circle, draw, inner sep=0pt,fill=black, minimum size=4pt] (u116) at (4.8,8) [label=left:] {};
\node[circle, draw, inner sep=0pt,fill=black, minimum size=4pt] (u117) at (5.6,8) [label=left:] {};
\node[circle, draw,fill=black, inner sep=0pt, minimum size=4pt] (u118) at (6.4,8) [label=left:] {};
\node[circle, draw,fill=black, inner sep=0pt, minimum size=4pt] (u119) at (7.2,8) [label=left:] {};
\node[circle, draw, fill=black,inner sep=0pt, minimum size=4pt] (u1110) at (8,8) [label=left:] {};
\node[circle, draw, fill=black,inner sep=0pt, minimum size=4pt] (u1111) at (8.8,8) [label=left:] {};
\node[circle, draw, fill=black, inner sep=0pt, minimum size=4pt] (u1112) at (9.6,8) [label=left:] {};
\node[circle, draw, fill=black, inner sep=0pt, minimum size=4pt] (u1113) at (10.4,8) [label=left:] {};

\node[circle, draw, inner sep=0pt,fill=black, minimum size=4pt] (u120) at (0,8.8) [label=left:] {};
\node[circle, draw, inner sep=0pt,fill=black, minimum size=4pt] (u121) at (.8,8.8) [label=left:] {};
\node[circle, draw, inner sep=0pt,fill=black, minimum size=4pt] (u122) at (1.6,8.8) [label=left:] {};
\node[circle, draw, inner sep=0pt, fill=black,minimum size=4pt] (u123) at (2.4,8.8) [label=left:] {};
\node[circle, draw, inner sep=0pt, fill=black,minimum size=4pt] (u124) at (3.2,8.8) [label=left:] {};
\node[circle, draw, inner sep=0pt, fill=black,minimum size=4pt] (u125) at (4,8.8) [label=left:] {};
\node[circle, draw, inner sep=0pt,fill=black, minimum size=4pt] (u126) at (4.8,8.8) [label=left:] {};
\node[circle, draw, inner sep=0pt,fill=black, minimum size=4pt] (u127) at (5.6,8.8) [label=left:] {};
\node[circle, draw,fill=black, inner sep=0pt, minimum size=4pt] (u128) at (6.4,8.8) [label=left:] {};
\node[circle, draw,fill=black, inner sep=0pt, minimum size=4pt] (u129) at (7.2,8.8) [label=left:] {};
\node[circle, draw, fill=black,inner sep=0pt, minimum size=4pt] (u1210) at (8,8.8) [label=left:] {};
\node[circle, draw, fill=black,inner sep=0pt, minimum size=4pt] (u1211) at (8.8,8.8) [label=left:] {};
\node[circle, draw, fill=black, inner sep=0pt, minimum size=4pt] (u1212) at (9.6,8.8) [label=left:] {};
\node[circle, draw, fill=black, inner sep=0pt, minimum size=4pt] (u1213) at (10.4,8.8) [label=left:] {};

\node[circle, draw, inner sep=0pt,fill=black, minimum size=4pt] (u130) at (0,9.6) [label=left:] {};
\node[circle, draw, inner sep=0pt,fill=black, minimum size=4pt] (u131) at (.8,9.6) [label=left:] {};
\node[circle, draw, inner sep=0pt,fill=black, minimum size=4pt] (u132) at (1.6,9.6) [label=left:] {};
\node[circle, draw, inner sep=0pt, fill=black,minimum size=4pt] (u133) at (2.4,9.6) [label=left:] {};
\node[circle, draw, inner sep=0pt, fill=black,minimum size=4pt] (u134) at (3.2,9.6) [label=left:] {};
\node[circle, draw, inner sep=0pt, fill=black,minimum size=4pt] (u135) at (4,9.6) [label=left:] {};
\node[circle, draw, inner sep=0pt,fill=black, minimum size=4pt] (u136) at (4.8,9.6) [label=left:] {};
\node[circle, draw, inner sep=0pt,fill=black, minimum size=4pt] (u137) at (5.6,9.6) [label=left:] {};
\node[circle, draw,fill=black, inner sep=0pt, minimum size=4pt] (u138) at (6.4,9.6) [label=left:] {};
\node[circle, draw,fill=black, inner sep=0pt, minimum size=4pt] (u139) at (7.2,9.6) [label=left:] {};
\node[circle, draw, fill=black,inner sep=0pt, minimum size=4pt] (u1310) at (8,9.6) [label=left:] {};
\node[circle, draw, fill=black,inner sep=0pt, minimum size=4pt] (u1311) at (8.8,9.6) [label=left:] {};
\node[circle, draw, fill=black, inner sep=0pt, minimum size=4pt] (u1312) at (9.6,9.6) [label=left:] {};
\node[circle, draw, fill=black, inner sep=0pt, minimum size=4pt] (u1313) at (10.4,9.6) [label=left:] {};

\path 
(u00) edge (u013)
(u10) edge (v113)
(u20) edge (u213)
(u30) edge (u313)
(u40) edge (u413)
(u50) edge (u513)
(u60) edge (u613)
(u70) edge (u713)
(u80) edge (u813)
(u90) edge (u913)
(u100) edge (u1013)
(u110) edge (u1113)
(u120) edge (u1213)
(u130) edge (u1313);

\path
(u00) edge (u130)
(u01) edge (u131)
(u02) edge (u132)
(u03) edge (u133)
(u04) edge (u134)
(u05) edge (u135)
(u06) edge (u136)
(u07) edge (u137)
(u08) edge (u138)
(u09) edge (u139)
(u010) edge (u1310)
(u011) edge (u1311)
(u012) edge (u1312)
(u013) edge (u1313);

\draw[black](u00)--(u1313);
\draw[black](u01)--(u1213);
\draw[black](u02)--(u1113);
\draw[black](u03)--(u1013);
\draw[black](u04)--(u913);
\draw[black](u05)--(u813);
\draw[black](u06)--(u713);
\draw[black](u07)--(u613);
\draw[black](u08)--(u513);
\draw[black](u09)--(u413);
\draw[black](u010)--(u313);
\draw[black](u011)--(u213);
\draw[black](u012)--(v113);

\draw[black](u00)--(u1313);
\draw[black](u10)--(u1312);
\draw[black](u20)--(u1311);
\draw[black](u30)--(u1310);
\draw[black](u40)--(u139);
\draw[black](u50)--(u138);
\draw[black](u60)--(u137);
\draw[black](u70)--(u136);
\draw[black](u80)--(u135);
\draw[black](u90)--(u134);
\draw[black](u100)--(u133);
\draw[black](u110)--(u132);
\draw[black](u120)--(u131);

\draw[rounded corners,color=blue, thick] (10.2, -.2) rectangle (10.6, 9.8) [label=left:]{}; 

\draw[rounded corners,color=blue, thick] (-.2, -1) rectangle (.2, 9.8) [label=left:]{}; 

\draw[rounded corners,color=blue, thick] (-.2, -.6) rectangle (9.8, -1) [label=left:]{};

\draw[rounded corners,color=blue, thick] (-.2, 9.8) rectangle (10.6, 9.4) [label=left:]{};

\draw[blue] (u013) .. controls (11.2,.5) .. (10.6,4.8);

\node[circle, draw, inner sep=0pt,color=red,fill=red, minimum size=6pt] (u22) at (1.6,.8) [label=left:] {};
\node[circle, draw, inner sep=0pt,color=red,fill=red,minimum size=6pt] (u26) at (4.8,.8) [label=left:] {};
\node[circle, draw, color=red,fill=red,inner sep=0pt, minimum size=6pt] (u210) at (8,.8) [label=left:] {};

\node[circle, draw, inner sep=0pt,color=red,fill=red,minimum size=6pt] (u62) at (1.6,4) [label=left:] {};
\node[circle, draw, inner sep=0pt,color=red,fill=red, minimum size=6pt] (u66) at (4.8,4) [label=left:] {};
\node[circle, draw, color=red,fill=red,inner sep=0pt, minimum size=6pt] (u610) at (8,4) [label=left:] {};

\node[circle, draw, inner sep=0pt,color=red,fill=red, minimum size=6pt] (u102) at (1.6,7.2) [label=left:] {};
\node[circle, draw, inner sep=0pt,color=red,fill=red, minimum size=6pt] (u106) at (4.8,7.2) [label=left:] {};
\node[circle, draw, color=red,fill=red,inner sep=0pt, minimum size=6pt] (u1010) at (8,7.2) [label=left:] {};

    \end{tikzpicture}
    \caption{Example of a triangulated grid $\Gamma_{14}$. Note that a blue edge denotes that the vertex is adjacent to all the vertices inside blue boundary. The set of vertices colored red forms $C'$.}
    \label{triangulated grid}
\end{figure}
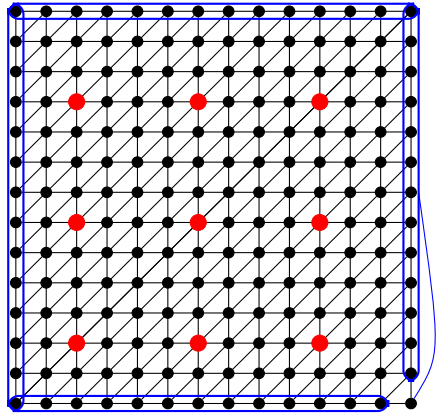

\noindent We now obtain the following result.
\begin{lemma}\label{grid}\rm
Let $G$ be a planar graph of minimum degree at least 2 which contains a triangulated grid $\Gamma_{10\sqrt{k}+4}$ as a minor obtained by only edge contraction operations. Then there always exists a locally minimal defensive alliance of size at least $k$.
\end{lemma}
\proof The proof of  Lemma \ref{grid} is similar to that of Lemma \ref{diameter}. Given that
$G$ can be transformed to $\Gamma_t$ after a sequence of edge contractions.
Suppose that the vertices of $\Gamma_{t}$ are labelled $(i,j)$ where $i, j \in [t]$.
Consider the set  $C'=\{(5x+2,5y+2)~|~ 0\leq x,y \leq \frac{t-4}{5}\} \subset V(\Gamma_t)$. See Figure \ref{triangulated grid} for an illustration; the vertices of $C'$ are shown in red.
Every (red) vertex  in $\Gamma_t$ is either an original vertex of $G$ or  obtained by contracting some edges of $G$. We obtain 
a set $C\subseteq V(G)$ from $C'$ as follows. For each (red) vertex $(i,j)$ in $C'$, 
if the red vertex is an original vertex then include it in $C$; if the red vertex is obtained by contracting some edges of $G$, 
then arbitrarily include in $C$ an endpoint  of one of the contracted edges.
Run Algorithm \ref{alg:cap} on $(G,V(G),C)$.
Let $D_{C}$ be a defensive alliance obtained at the end of the {\bf for} loop of lines 1-5  in Algorithm \ref{alg:cap}.
Note that $D_{C}$ may not be a connected defensive alliance. Let $S_{1},S_{2},\ldots,S_{p}$ be connected components of $D_{C}$ such that $\bigcup\limits_{i=1}^{p} S_{i} = D_{C}$. 
This implies that $D_{C}$ will generate a partition of $C$ into $C_{1},C_{2},\ldots,C_{p}$ such that $C_{i} \subseteq S_{i}$.

\begin{claim}
Line 6 of Algorithm \ref{alg:cap} runs Algorithm \ref{algo1} on  $(G,S_{i})$ and
outputs a locally minimal defensive alliance $S'_{i}$ such that $|S'_{i}| \geq \max\big\{2,|C_{i}|-1\big\}$ for all $1\leq i \leq p$.
\end{claim}

\noindent{\it Proof of Claim:} Due to Observation \ref{obs1}, we have $|S'_{i}|\geq 2$.
Therefore let  us focus on $C_i$'s such that $|C_i|\geq 4$. Let $C_{i}= \{c_{1},c_{2},\ldots,c_{q}\}$. Then $d_G(c_{s},c_{t}) \geq 4$ for all $1\leq s,t \leq q-1$ and $s\neq t$ due to choice of $C$.
Note that as $G[S_{i}]$ is connected, it implies that $G[S_{i}]$ contains at least $q-1$ crucial vertices in the defensive alliance $S_{i}$. 
This is because any path between $c_{s}$ and $c_{t}$ in $G[S_{i}]$ contains a vertex $u$ such that $d(u,C_{i})\geq 2$ for all $1\leq i \leq q-1$. Due to Lemma \ref{cruciallemma}, we have $|S'_{i}| \geq q-1$. This completes the proof of the lemma.\\

\noindent If we set $t=10\sqrt{k}+4$ then $|C|\geq 2k$. It implies that $\bigcup\limits_{i=1}^{p} {S'_i}$ is a locally minimal defensive alliance of size at least $k$. This completes the proof of the lemma. \qed\\\\

\noindent To design a dynamic programming algorithm on a given tree decomposition of input graph, we use the following theorem.

\begin{theorem}\label{tree}\cite{10.1007/978-3-030-67899-9_11}
Given an $n$-vertex graph $G$ and its nice tree decomposition $T$ of width at most $w$, the 
size of a maximum locally minimal defensive alliance of $G$ can be computed in time $8^w \Delta^{\mathcal{O}(2^{w})}$.
\end{theorem}

\noindent Finally to design an FPT algorithm when parameterized by the solution size $k$, we prove the following lemma.

\begin{lemma}\label{planar}\rm
Let $G$ be a planar graph with minimum degree at least 2. If $\Delta(G) \geq k^{\mathcal{O}(k)}$ for some sufficiently large constant then there always exists a locally minimal defensive alliance of size at least $k$.
\end{lemma}

\proof Suppose $d(u_0) \geq k^{\mathcal{O}(k)}$.
Run BFS$(G,u_0)$ to obtain the BFS tree
$T$ rooted at $u_0$ and to get level for each vertex in $G$. Run Algorithm \ref{alg:cap} on $(G,V(G),u_{0})$ and suppose it outputs $D_0$. 
If $u_0$ is in $D_0$ then at least $\frac{d(u_0)}{2}>k$ neighbours  of $u_0$ are in $D_0$ for its protection. Therefore $D_0$ is a locally minimal defensive alliance of size at least $k$. 
Consider the case where $u_0$ is not in $D_0$. 
Let $D'_0$ be the defensive alliance obtained at the end of the {\bf for} loop of 
lines 1-5 in Algorithm \ref{alg:cap} and 
all neighbours of $u_0$ in $D'_0$ are overprotected. 
This is why $u_0$ is deleted from $D'_0$. 
As we have discussed before, the vertices in  $ D'_0\cap L_2$ are crucial. 
Therefore if  $ D'_0\cap L_2$ contains more than $k-1$ vertices or any of them has degree more than $2k$ then we have a locally minimal defensive alliance
of size at least $k$. So we assume $ |D'_0\cap L_2|\leq k-1$ and every vertex 
in $D'_0\cap L_2$ has degree less than $2k$.
Now we consider the defensive alliance $D''_0=D'_0\setminus \{u_0\}$. Clearly
$D''_0 \cap L_1$ contains at least $k^{\mathcal{O}(k)}$ vertices.
Let us consider the graph $G[D_{0}'']$.  If $G[D_{0}'']$ contains 
$\frac{k}{2}$ or more connected components of size at least two, 
then by Observation \ref{obs1} and \ref{Union1}, $G$ has a locally minimal defensive alliance of size 
at least $k$. If the diameter of $G[D_{0}'']$  is more than $8k$ then by Lemma  \ref{diameter}, $G$ has a locally minimal defensive alliance of size at least $k$. Thus we assume that 
$G[D_{0}'']$ has at most $\frac{k}{2}-1$ components and the diameter of each component
of $G[D_{0}'']$ is less than $8k$. Therefore if the maximum degree of $G[D''_0]$ is $f_{1}(k)$ then we know that it contains at most $\mathcal{O}(kf_{1}(k)^{k})$ vertices. We also know that $G[D''_0]$ contains at least $k^{\mathcal{O}(k)}$ vertices. Therefore, we get $k^{c_1k}\leq c_2 k \cdot f_1(k)^{k}$ where $c_1$ and $c_2$ are two positive real numbers. This implies
$$ f_1(k)\geq \frac{1}{c_2}k^{c_1-1}.$$
Take $c_1=3$. Therefore, we can assume that there exists  a vertex $u_1$ in  $L_1 \cap D''_0$ with degree at least $\mathcal{O}(k^{2})$ with sufficiently large constant. 
Again run Algorithm \ref{alg:cap} on $(G,D''_0, \{u_1\})$ and suppose it outputs $D_{1}$. If $u_1 \in D_{1}$ then clearly $D_{1}$ is a locally minimal defensive alliance of size at least $k$. 
Consider the case where $u_1$ is not in $D_{1}$.  
Let $D'_{1}$ be the defensive alliance obtained at line 5 of Algorithm 2. Now consider the defensive alliance $D''_{1}=D'_{1}\setminus \{u_1\}$.

\begin{claim}
The defensive alliance $D''_{1}$ either contains at least  $\frac{k}{2}$ (locally minimal) defensive alliances $S_{1},\ldots,S_{\frac{k}{2}}$ such that the closed neighbourhood of $S_i$ does not intersect  $S_j$ 
 for all $1\leq i,j \leq \frac{k}{2}$, $i\neq j$ or the diameter of one of the connected components of $D''_{1}$ is more than $8k$.
\end{claim}

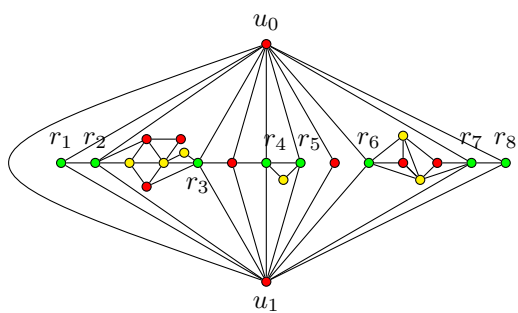
\begin{figure}
    \centering
  \begin{tikzpicture}[scale=0.45]
  
\node[circle,draw,fill=red, inner sep=0 pt, minimum size=0.12cm]	(u) at (0,1) [label=above:$u_0$]{}; 
\node[circle,draw,fill=red, inner sep=0 pt, minimum size=0.12cm]	(u0) at (0,-6) [label=below:$u_1$]{};     
    
\node[circle,draw,fill=green, inner sep=0 pt, minimum size=0.12cm]	(r1) at (-6,-2.5) [label=above:$r_{1}$]{};    
\node[circle,draw,fill=green, inner sep=0 pt, minimum size=0.12cm]	(r2) at (-5,-2.5) [label=above:$r_{2}$]{};
\node[circle,draw,fill=yellow, inner sep=0 pt, minimum size=0.12cm]	(r3) at (-4,-2.5) [label=above:]{};   

\node[circle,draw,fill=red, inner sep=0 pt, minimum size=0.12cm]	(e1) at (-3.5,-1.8) [label=above:]{};   
\node[circle,draw,fill=red, inner sep=0 pt, minimum size=0.12cm]	(e2) at (-3.5,-3.2) [label=above:]{};   
\node[circle,draw,fill=red, inner sep=0 pt, minimum size=0.12cm]	(e3) at (-2.5,-1.8) [label=above:]{};  
\node[circle,draw,fill=yellow, inner sep=0 pt, minimum size=0.12cm]	(e4) at (-2.4,-2.2) [label=above:]{};  

\node[circle,draw,fill=yellow, inner sep=0 pt, minimum size=0.12cm]	(r4) at (-3,-2.5) [label=above:]{};    
\node[circle,draw,fill=green, inner sep=0 pt, minimum size=0.12cm]	(r5) at (-2,-2.5) [label=below:$r_{3}$]{};    
\node[circle,draw,fill=red, inner sep=0 pt, minimum size=0.12cm]	(r6) at (-1,-2.5) [label=above left:]{};   
\node[circle,draw,fill=green, inner sep=0 pt, minimum size=0.12cm]	(r7) at (0,-2.5) []{};    

\node[circle,draw,fill=yellow, inner sep=0 pt, minimum size=0.12cm]	(e7) at (0.5,-3) [label=above:]{};

\node[circle,draw,fill=green, inner sep=0 pt, minimum size=0.12cm]	(r8) at (1,-2.5) [label=above left:$r_{4}$]{};    
\node[circle,draw,fill=red, inner sep=0 pt, minimum size=0.12cm]	(r9) at (2,-2.5) [label=above left:$r_{5}$]{};    
\node[circle,draw,fill=green, inner sep=0 pt, minimum size=0.12cm]	(r11) at (3,-2.5) [label=above:$r_{6}$]{};    
\node[circle,draw,fill=red, inner sep=0 pt, minimum size=0.12cm]	(r12) at (4,-2.5) [label=above:]{};

\node[circle,draw,fill=yellow, inner sep=0 pt, minimum size=0.12cm]	(e5) at (4,-1.7) [label=above:]{}; 
\node[circle,draw,fill=yellow, inner sep=0 pt, minimum size=0.12cm]	(e6) at (4.5,-3) [label=above:]{}; 

\node[circle,draw,fill=red, inner sep=0 pt, minimum size=0.12cm]	(r13) at (5,-2.5) [label=above:]{};    
\node[circle,draw,fill=green, inner sep=0 pt, minimum size=0.12cm]	(r14) at (6,-2.5) [label=above:$r_{7}$]{};  
\node[circle,draw,fill=green, inner sep=0 pt, minimum size=0.12cm]	(r15) at (7,-2.5) [label=above:$r_{8}$]{}; 

\draw(u)--(r1);
\draw(u0)--(r1);    
\draw(u)--(r2);
\draw(u0)--(r2);  
\draw(r2)--(r1);
\draw(r3)--(r2);
\draw(e1)--(r2);  
\draw(r3)--(e1);
\draw(r4)--(e1);  
\draw(r3)--(r4);
\draw(r3)--(e2);  
\draw(r4)--(e2);
\draw(r5)--(e2);  
\draw(u)--(r5);
\draw(e3)--(e1);
\draw(e3)--(r4);
\draw(r4)--(r5);
\draw(u0)--(r5);
\draw(u)--(r6);
\draw(u0)--(r6);
\draw(u)--(r7);
\draw(u0)--(r7);
\draw(u)--(r8);
\draw(u0)--(r8);
\draw(r7)--(r6);
\draw(r6)--(r5);
\draw(e4)--(r4);
\draw(e4)--(r5);
\draw(r7)--(r8);
\draw(u)--(r9);
\draw(u0)--(r9);
\draw(u)--(r11);
\draw(u0)--(r11);
\draw(u)--(r14);
\draw(u0)--(r14);
\draw(u)--(r15);
\draw(u0)--(r15);
\draw(e6)--(e5);
\draw(r11)--(r12);
\draw(e5)--(r12);
\draw(e6)--(r12);
\draw(r11)--(e5);
\draw(r13)--(r14);
\draw(e6)--(r12);
\draw(e6)--(r13);
\draw(e6)--(r14);
\draw(r15)--(r14);
\draw(e6)--(r11);
\draw(e5)--(r13);
\draw(e7)--(r7);
\draw(e7)--(r8);

\draw (u) .. controls (-10,-2.5) ..  (u0);

  \end{tikzpicture}
    \caption{A planar drawing of the vertices in $R \cup \{u_0,u_1\}$. The vertices colored green and yellow are inside the defensive alliance and vertices colored red are outside the defensive alliance.}
    \label{plnar graphs}
\end{figure}
 
\noindent{\it Proof of Claim:} First, we observe that $D_{1}^{\prime \prime}$ contains  at least $\mathcal{O}(k^{2})$ vertices which are adjacent to both $u_0$ and $u_1$. 
Let $R=N(u_0)\cap N(u_1) \cap D_{1}^{\prime \prime}$ and let  $ r_{1},r_{2},\ldots,r_{c.k^{2}}$ (where
 $c$ is a positive real number) be  an ordering of the vertices of $R$ so that we have a planar drawing of $G[R\cup\{u_0,u_1\}]$. 
As $u_0,u_1 \not\in D_{1}^{\prime \prime}$ and $G$ is planar, we can say that a 
path from $r_{i}$ to $r_{j}$ in $G[D_{1}^{\prime \prime}]$ must pass through all the vertices of $\{r_{i+1},r_{i+2}\ldots,r_{j-1}\}$ for all $1\leq i < j \leq c.k^{2}$. 
See Figure \ref{plnar graphs} for an illustration of a planar drawing of the vertices of $R\cup \{u_0,u_1\}$. The vertices of $R$ are shown in green, the vertices of 
$D_{1}^{\prime \prime} \setminus R$ are shown in yellow where as the 
vertices outside $D_{1}^{\prime \prime}$ are shown in red. 
 Let $\max_i$ be the largest index such that 
there is a path from $r_{\max_{i-1}+1}$ to $r_{\max_i}$ in $G[D''_1]$, that is, using the yellow and green vertices only. Set $\max_0=0$.
For $i=1$, there is a path from $r_1$ to $r_{\max_1}$ in $G[D''_1]$ and the length 
of this path is at least $\max_1$.
If $\max_1\geq 8k$ then  the diameter of this  component 
is more than $8k$. Therefore, we assume that $\max_1< 8k$. The connected 
component in $G[D_{1}^{\prime \prime}]$ containing all the vertices of $\{r_{1},r_{2},\ldots,r_{\max_{1}}\}$ is denoted by $S_{1}$. 
  Clearly  $S_{1}$ is a defensive alliance. 
  Note that $S_{1}$ does not contain any vertex from $\{r_{\max_{1}+1},\ldots,r_{c.k^{2}}\}$.
  Next, for $i=2$, there is a path from $r_{\max_1+ 1}$ to  $r_{\max_{2}}$  
  in $G[D''_1]$. In Figure \ref{plnar graphs} we have 
  $\max_1=3$, $\max_2=5$ and $\max_3=8$. As seen before, we must have $\max_{2}-\max_{1}-1\leq  8k$, otherwise the diameter of this component is more than $8k$. 
The connected  component in $G[D_{1}^{\prime \prime}]$ containing all the vertices of $\{r_{\max_{1}+1},\ldots,r_{\max_{2}}\}$ is denoted by $S_{2}$. Clearly  $S_{2}$ is also a defensive alliance. 
Repeat this process $\frac{k}{2}$ times.  We get either at least $\frac{k}{2}$ defensive alliances $S_{1},\ldots,S_{\frac{k}{2}}$ such that $N[S_{i}] \cap S_{j} = \emptyset$ for all $1\leq i,j \leq k$, $i\neq j$ or a component  in $G[D''_{1}]$ with diameter more than $8k$.  By Observation \ref{Union1} and Lemma \ref{diameter}, we get a locally minimal defensive alliance of size at least $k$ in $G$. \qed\\\\

\begin{theorem} The
{\sc Locally Minimal Defensive Alliance} problem on the planar graphs
 of minimum degree at least 2, admits an FPT algorithm with running time  $\mathcal{O}^{*}( k^{2^{\mathcal{O}(\sqrt{k})}})$.
\end{theorem}

\proof By Lemma \ref{planar}, if $\Delta(G) \geq k^{\mathcal{O}(k)}$ then $(G,k)$ is a 
yes-instance. 
Therefore we can assume that $\Delta(G) < k^{\mathcal{O}(k)}$. 
By Theorem \ref{planargrid}, if the treewidth of $G$ is more than 
$\mathcal{O}(\sqrt{k})$ for some sufficiently large constant then  $G$ contains $\Gamma_{10\sqrt{k}+4}$
as a contraction. In this case, by Lemma \ref{grid},  $(G,k)$ is a 
yes-instance. 
Therefore, we  also assume that the treewidth of $G$ is at most $\mathcal{O}(\sqrt{k})$. 
Now we run the algorithm of Theorem \ref{tree} on graphs with  $\Delta(G) < k^{\mathcal{O}(k)}$ and $tw(G)\leq \mathcal{O}(\sqrt{k})$ to solve the problem in time
$\mathcal{O}^{*}(k^{2^{\mathcal{O}(\sqrt{k})}})$. \qed\\

\section{Conclusion and Open Problems}

We proved that {\sc Locally Minimal Defensive Alliance} admits an FPT algorithm on general graphs with minimum degree at least 2, when parameterized only by the solution size. 
When the problem is restricted to $C_3$-free and $C_4$-free graphs of minimum degree at least 2, we get a kernel of size $k^{\mathcal{O}(k)}$. We also provide an FPT algorithm with running time  $k^{2^{\mathcal{O}(\sqrt{k})}}$ on planar graphs  with minimum degree at least 2.
In the above discussion, we have mostly dealt  with graphs of minimum degree at least $2$. 
We observe that we can remove this condition while dealing with locally minimal strong defensive alliance. 
We list some problems emerge from the results here:
 \begin{enumerate}
 
 \item It remains open whether {\sc Locally Minimal Defensive Alliance} on general graphs, without any restriction on minimum degree, belong to FPT when parameterized  by the
 solution size alone.
 
 \item Whether {\sc Connected Locally Minimal Defensive Alliance} on general graphs belong to FPT when parameterized only by the
 solution size.
 
 \item Can we improve the size of the kernel obtained for {\sc Locally Minimal Defensive Alliance} on general graphs or obtain efficient FPT algorithms on  special graph classes?
 \item  It may be interesting to study if our ideas can be useful to see whether {\sc Globally
 Minimal Defensive Alliance}  belong to  FPT when parameterized by the
 solution size. 

 \end{enumerate}

\noindent We also propose the following graph theoretic problem.
If we can design a polynomial-time algorithm to determine whether there exists a locally minimal defensive alliance containing a given vertex $v$ or not then this will  drastically improve the size of kernel on general graphs.  It is already proved in \cite{https://doi.org/10.48550/arxiv.2202.02010} that there is no polynomial algorithm to determine whether there exists a globally minimal defensive alliance containing a given vertex $v$ or not.

\bibliographystyle{abbrv}
\bibliography{bibliography}
\newpage


\end{document}